\documentclass[aps,pre,reprint,groupedaddress,showpacs,onecolumn]{revtex4-1}

\usepackage{amsmath}
\usepackage{graphicx}
\usepackage{float}
\usepackage{algpseudocode}
\usepackage{color}
\usepackage{hyperref}
\hypersetup{pdftitle=The SAT-UNSAT transition in the adversarial SAT problem, pdfauthor={Marco Bardoscia, Daniel Nagaj, Antonello Scardicchio}}

\begin{document}

%hack for floating algorithms
\floatstyle{ruled}
\newfloat{algorithm}{thp}{alg}
\floatname{algorithm}{Algorithm }

\title{The SAT-UNSAT transition in the adversarial SAT problem}

\author{Marco Bardoscia} 
\email{marco.bardoscia@ictp.it}
\affiliation{Abdus Salam International Centre for Theoretical Physics, Strada Costiera 11, 34151 Trieste, Italy}
\author{Daniel Nagaj}
\email{dnagaj@gmail.com}
\affiliation{Dept.\ of Physics, University of Vienna, Boltzmanngasse 5, 1090 Vienna, Austria}
\affiliation{Institute of Physics, Slovak Academy of Sciences, D\'{u}bravsk\'{a} cesta 9, 845 11 Bratislava, Slovakia}
\author{Antonello Scardicchio}
\email{ascardic@ictp.it}
\affiliation{Abdus Salam International Centre for Theoretical Physics, Strada Costiera 11, 34151 Trieste, Italy}
\affiliation{INFN, Section of Trieste, Strada Costiera 11, 34151 Trieste, Italy}

\date{\today}

\begin{abstract}
Adversarial SAT (AdSAT) is a generalization of the satisfiability (SAT) problem in which two players try to make a boolean formula true (resp.\ false) by controlling their respective sets of variables.
AdSAT belongs to a higher complexity class in the polynomial hierarchy than SAT and therefore the nature of the critical region and the transition are not easily paralleled to those of SAT and worth of independent study. AdSAT also provides an upper bound for the transition threshold of the quantum satisfiability problem (QSAT). We present a complete algorithm for AdSAT, show that 2-AdSAT is in $\mathbf{P}$, and then study two stochastic algorithms (simulated annealing and its improved variant) and compare their performances in detail for 3-AdSAT. Varying the density of clauses $\alpha$ we find a sharp SAT-UNSAT transition at a critical value whose upper bound is $\alpha_c \lesssim 1.5$, thus providing a much stricter upper bound for the QSAT transition than those previously found.
\end{abstract}

%\keywords{}

\maketitle

\section{Introduction} 
\label{sec:intro}

The study of random ensembles of decision problems has grown into a fertile field of investigation where the methods of statistical Physics have found applications to the theory (and practice) of hard combinatorial problems. This resulted in a wealth of intuition on the nature of the typical complexity of hard decision problems and in a new, efficient family of algorithms \cite{mezard2002analytic, friedgut1999sharp}. One such problem is a random ensemble of satisfiability (in short SAT), where boolean formulas are generated in a random way and tested for a solution. If the formula is restricted to be of the form of a conjunction of an arbitrary number of clauses, and each clause is the logical disjunction of $K$ variables, the problem is denoted by $K$-SAT. The ensemble is determined once the number of clauses per variable is fixed. As this ratio is increased the formulas go from being typically satisfiable to being typically unsatisfiable \cite{kirkpatrick1994critical,monasson1999determining,friedgut1999sharp}. This is the SAT-UNSAT phase transition.

Recent progress in the study of \emph{quantum} decision problems \cite{kempe2006complexity} lead to the definition of the quantum generalisation of $K$-SAT (we call it $K$-QSAT) \cite{bravyi2006efficient}. This problem is proven to be $\mathbf{QMA}_1$ complete for $K\geq 3$ \cite{gosset2013quantum}, with $\mathbf{QMA}_1$ the quantum analog of $\mathbf{NP}$\footnote{To be more precise, $\mathbf{QMA}_1$ is the quantum analog of $\mathbf{MA}_1$, where the verification procedure allows false instances to be accepted with a small probability.}. A random ensemble of $K$-QSAT was introduced and studied in \cite{laumann2010random, laumann2010product,laumann2012statistical}, where it has been shown to have a SAT-UNSAT phase transition. Moreover, it has a phase with product-state solutions. The quantum version of the Lov\'{a}sz local lemma in \cite{ambainis2012quantum} provides a lower bound for the SAT-UNSAT transition which implies that sufficiently close to the phase transition the solutions of the problem must be entangled (this is proven for sufficiently large $K$ and believed true for all $K \geq 3$).

Following a theorem in \cite{laumann2010random}, an upper bound on the quantum SAT-UNSAT threshold can be found in terms of \emph{the most frustrated classical formula} on a given hyper-graph (as it will be explained in the following, the graph is determined by the membership relations between variables and clauses). This problem is interesting for at least two other reasons. First, it is a random problem of satisfiability formulae with more than one existential quantifier (it actually has two quantifiers) and second, it is a problem of extreme-value statistics on the familiar random ensemble of $K$-SAT formulae.

The problem was named adversarial SAT (AdSAT) in \cite{castellana-zdeborova}, where it has been studied by means of belief and survey propagation. There it is claimed that the problem has a SAT-UNSAT transition for $K=3$ at a clauses/variables ratio $\alpha=3.39\pm0.01$, (compare with the familiar SAT-UNSAT transition of $K$-SAT at $\alpha=4.27$). If this result is correct then AdSAT improves only marginally on the upper bound for the transition in QSAT. In fact, an upper bound is found in \cite{bravyi2009bounds} at $\alpha=3.59$. However, from the numerics in \cite{laumann2010random} the threshold for 3-QSAT is closer to $\alpha=1.0$ than to any one of these numbers, signaling that the physics of QSAT is not captured well by these approximations. 

In this paper we numerically investigate the random ensemble of AdSAT problems. Since the problem is quite resilient to numerical analysis, we present two heuristic algorithms and study their performances. By these means, we are able to study the crossover of the SAT-UNSAT transition at varying $N$. For the largest system size systematically explored ($N=15$) we observe an $\alpha_c=1.6$  and a clear tendency of $\alpha_c$ to decrease with increasing $N$. We support this picture with investigations of considerably larger system sizes ($N=100$) where we can assert that $\alpha_c<2.70$. To reconcile these numbers with the analytical results of \cite{castellana-zdeborova}, one concludes that the $1/N$ and $1/N^2$ corrections required here are really large (the coefficient of $1/N$ should be of $O(10^2)$).

The paper is organized as follows. In Sec.\ \ref{sec:preliminaries} we formally introduce the AdSAT problem and we briefly discuss its importance from the perspective of complexity theory, showing an efficient approach to 2-AdSAT in Appendix \ref{sec:adsat2}. In Sec.\ \ref{sec:complete} we present a simple complete algorithm to solve the AdSAT problem. Due to the complexity of the problem we resort to a stochastic algorithm based on simulated annealing, which is introduced in Sec.\ \ref{sec:annealing}, and in Sec.\ \ref{sec:improved} we discuss its much improved variant, investigating the SAT-UNSAT transition for 3-AdSAT. Our conclusions are summarized in Sec.\ \ref{sec:conclusions}.

%%%%%%%%%%%%%%%%%%%%%%%%%%%%%%%%%%%%%%%%%%%%%%%%%%%%%%%%%%%%%%%%%%%%%%%%%%%%%%%%%%%%%%%%%%%%%%%%%%%%

\section{Preliminaries}
\label{sec:preliminaries}

An AdSAT formula $\phi_G$ is a boolean function of $N$ boolean variables $\{x_i\}_{i=1,\ldots,N}$, which will be referred to as \emph{bits} and $KM$ boolean variables $\{J_{aj}\}_{\substack{a=1,\ldots,M\\j=1,\ldots,K}}$, which we call \emph{negations}. It can be written in terms of $M$ clauses $\{C_a\}_{a=1,\ldots,M}$, each clause having the structure
\begin{equation} \label{eq:clause}
	C_a = \bigvee_{j=1}^K x_{aj} \oplus J_{aj} \, ,
\end{equation}
where $x_{aj}$ is the bit (taken from $\{x_i\}_{i=1,\ldots,N}$) occupying the $j$th place in the $a$th clause. 
The negation $J_{aj}\in\{0,1\}$ decides whether the variable $x_{aj}$ or its negation $\overline{x}_{aj}$ appears in the final form of the clause $C_a$. For example, the clause
\begin{equation}
	C_b = (x_3\oplus 0) \vee (x_5\oplus 1) \vee (x_8\oplus 0) =  x_3 \vee \overline{x}_5 \vee x_8 \, ,
\end{equation}
has $x_{b1}=x_3,x_{b2}=x_5,x_{b3}=x_8$ and $J_{b3}=0,J_{b5}=1,J_{b8}=0$.
The AdSAT formula is then a conjunction of all the clauses:
\begin{equation} \label{eq:formula}
	\phi_G(x,\mathcal{J}) = \bigwedge_{a=1}^M C_a \, ,
\end{equation}
where $\mathcal{J} = \{J_{aj}\}$. Each bit can appear at most once in each clause, and each clause can appear at most once in the formula. 

\begin{figure}
	\centering
	\includegraphics[width=2.5cm]{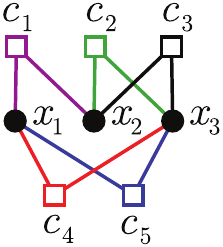}
	\caption{A graph of an AdSAT instance with $K=2$, $M=5$ and $N=3$. Each square corresponds to a disjunctive clause (\ref{eq:clause}) involving the bits (circles) or their negations. AdSAT is a game in which an adversary fixes the negations $\mathcal{J}$ on the edges, and challenges a player to set the bits $x$ so that the formula (\ref{eq:formula}) is true. When the adversary has a winning strategy, the AdSAT formula $\phi_G$ and its graph $G$ is called UNSAT.}
	\label{fig:hypergraph}
\end{figure}

An AdSAT formula $\phi_G$ can be conveniently represented by a bipartite graph $G$ as in Fig.\ \ref{fig:hypergraph} in which the nodes are of two kinds: variable nodes, and clause nodes. A variable node is associated to each bit, and a clause node to each clause. If the bit $x_i$ appears in the clause $C_a$ there is an edge between the corresponding variable and clause nodes. The total number of edges is $KM$, and we can imagine a negation as an attribute of each edge. 

An AdSAT formula is said to be SAT if:
\begin{equation} \label{eq:adsat-sat}
	\phi_G \in L_{\mathrm{SAT}}	\, \Leftrightarrow \, \forall \mathcal{J},  \exists x : \phi_G(x,\mathcal{J}) = 1 \, ,
\end{equation}
i.e.\ for any choice of the negations $\mathcal{J}$, there is a way to assign the bits $x$ so that the formula $\phi_G(x,\mathcal{J})$ is true. On the other hand, we call an AdSAT formula UNSAT if
\begin{equation}\label{eq:adsat-unsat}
	\phi_G \in L_{\mathrm{UNSAT}} \,  \Leftrightarrow \,  \exists \mathcal{J},  \forall x : \phi_G(x,\mathcal{J}) = 0 \, ,
\end{equation}
i.e.\ there exists an assignment of the negations $\mathcal{J}$ such that no choice of the bits $x$ can make the formula $\phi_G(x,\mathcal{J})$ true.
Since the property of being either SAT or UNSAT depends on the geometric structure of the clauses (i.e.\ on the bipartite graph $G$ associated to the formula), we could also say that $G$ is either SAT or UNSAT. Once the configuration of negations in an AdSAT formula is fixed, we are left with a standard $K$-SAT formula that will be denoted with $\phi_G(\cdot,\mathcal{J})$. In the following we will use the expressions SAT and UNSAT with a different meaning that will depend on the context with reference both to an AdSAT formula and to a standard $K$-SAT formula. In the former case their meaning is defined by (\ref{eq:adsat-sat}) and (\ref{eq:adsat-unsat}). In the latter case a formula is SAT if an assignment of bits exists such that the formula is true; vice versa the formula is UNSAT if it is false for all the possible assignments of bits.

The name ``adversarial'' SAT comes naturally when we view the problem as a game between two players: a positive player controls the $N$ bits, while the adversary player controls the $KM$ negations. First, the adversary chooses the negations.
The positive player wins if he is now able to find a configuration of bits such that the formula $\phi_G(x,\mathcal{J})$ is true.
Conversely, the adversarial player wins if he is able to find a configuration of negations such that no matter which configuration of bits the positive player chooses, the formula $\phi_G(x,\mathcal{J})$ is false. 
For a fixed $\mathcal{J}$, the positive player has $2^N$ possible configurations to choose from and he is facing a standard $K$-SAT problem (he wins if he is able to prove that the formula $\phi_G(\cdot,\mathcal{J})$ is SAT). On the other hand, the adversarial player has $2^{KM}$ configurations available in principle. Still, at least for $N$ negations his choice is straightforward. Suppose for example that $x_5$ appears for the first time in the second clause in the third position ($x_{23} = x_5$), then without loss of generality $J_{23}$ (corresponding to the first appearance of bit $x_5$) can be immediately set to zero. We can do this for every bit, fixing $N$ of the negations, so that the number of effective configurations for the second player is $2^{KM - N}$.

We study the typical behavior of the random AdSAT problem, i.e.\ the problem in which the graph associated with the formula is a random graph. In particular, we will focus on the ensemble in which $N$, $M$ and $K$ are fixed and each position in a clause can be occupied with equal probability by each bit, given the aforementioned constraints. Extensive numerical \cite{kirkpatrick1994critical,monasson1999determining} and analytical \cite{mezard2002analytic,monasson1999determining,friedgut1999sharp} evidence suggests that for the random $K$-SAT problem \cite{hartmann-weigt} in the limit $N, M \rightarrow \infty$ with finite $\alpha = M/N$, there is a critical value $\alpha_c^{\mathrm{SAT}}$ such that for $\alpha < \alpha_c^{\mathrm{SAT}}$ a formula is almost surely satisfiable, while for $\alpha > \alpha_c^{\mathrm{SAT}}$ it is almost surely unsatisfiable. If a similar critical value existed also for AdSAT, it would clearly be $\alpha_c^{\mathrm{AdSAT}} < \alpha_c^{\mathrm{SAT}}$. In fact, as already noted, a $K$-SAT formula is simply an AdSAT formula with frozen negations and if it is not possible to make an AdSAT formula UNSAT it means that all the $K$-SAT formulas with frozen negations must be SAT. It has been shown in \cite{laumann2010random} that $\alpha_c^{K\mathrm{-QSAT}} \leq \alpha_c^{\mathrm{AdSAT}}$.

Beyond the connection with QSAT, AdSAT is also relevant on its own, and should find its place in the perspective of complexity theory \cite{arora-barak}. Complexity theory classifies decision problems according to their algorithmic difficulty using a whole hierarchy of classes. Roughly speaking, a problem is considered ``easy'' (and it is said to belong to the class $\mathbf{P}$) if, in the worst case, it can be solved in a time scaling polynomially with the size of the problem. On the other hand we define the class $\mathbf{NP}$, which contains the problems for which it is possible to verify in a polynomial time if a candidate assignment is a solution of the problem. The $K$-SAT problem is in $\mathbf{NP}$ (we can directly check if an assignment of bits makes a boolean expression true) while it is not believed to be in $\mathbf{P}$ as this would imply (with $K\geq 3$) that $\mathbf{P}=\mathbf{NP}$. 

A natural increase in the complexity of the problem gives us the class $\mathbf{\Sigma}_2^p$ ($\mathbf{NP}$ with a $\mathbf{coNP}$ oracle). A problem is in $\mathbf{\Sigma}_2^p$ if its variables can be divided into two groups $u$ and $v$, so that given a candidate assignment $u^*$, for all the possible assignments of $v$ it is possible to verify in a polynomial time that $u^*$ is actually a solution of the problem. Since there are $2^N$ possible assignments of $v$ (hence not scaling polynomially with the size of the problem) we can only assert that $\mathbf{P} \subseteq \mathbf{NP} \subseteq \mathbf{\Sigma}_2^p$. Establishing if those inclusions are proper or not is probably the most important problem of contemporary computer science.

AdSAT is in $\mathbf{\Sigma}_2^p$ because given a candidate adversarial assignment of negations (this corresponds to $u^*$ in the definition of $\mathbf{\Sigma}_2^p$), the problem reduces to certifying that the formula $\phi_G(\cdot,\mathcal{J})$ is UNSAT. We can do this with a $\mathbf{coNP}$ oracle, which implies that AdSAT belongs to $\mathbf{\Sigma}_2^p$. It would be of great importance to understand if AdSAT, despite the appearance, is in $\mathbf{NP}$ or not.\footnote{This might appear hopeless. However, AdSAT is a graph property, so it is imaginable that it has a compressed testing procedure. Moreover, our attempts to show that AdSAT is $\mathbf{\Sigma}_2^p$-complete were not successful.} 
Note that 2-SAT is in $\mathbf{P}$. Therefore, 2-AdSAT (AdSAT for $K=2$) is naturally contained in $\mathbf{NP}$, as it has a short verification procedure. The witness is an adversarial assignment $\mathcal{J}^*$ of the negations. Given these, we can efficiently assert that the 2-SAT formula $\phi_G(\cdot,\mathcal{J}^*)$ is UNSAT, proving that the original adversarial SAT formula $\phi_G$ is UNSAT. However, the 2-AdSAT problem is even easier. We can calculate whether a 2-AdSAT formula is SAT/UNSAT by investigating the properties of the graph $G$, as we show in Appendix \ref{sec:adsat2}. For the rest of the paper we will focus on the case $K=3$.

%%%%%%%%%%%%%%%%%%%%%%%%%%%%%%%%%%%%%%%%%%%%%%%%%%%%%%%%%%%%%%%%%%%%%%%%%%%%%%%%%%%%%%%%%%%%%%%%%%%%

\section{A complete algorithm} 
\label{sec:complete}

A complete algorithm for AdSAT must be able to determine with certainty if a given AdSAT formula $\phi_G$ is SAT or UNSAT. A possible straightforward algorithm consists in solving the $3$-SAT formulae $\phi_G(\cdot, \mathcal{J})$ for all the possible configurations of negations $\mathcal{J}$. As soon as one configuration of negations such that $\phi_G(\cdot, \mathcal{J})$ is UNSAT is found, then the algorithm can stop as $\phi_G$ is also UNSAT. On the contrary, if $\phi_G(\cdot, \mathcal{J})$ is SAT on all the possible configurations of negations, then $\phi_G$ is SAT. In order to solve the $3$-SAT formulae one must use a complete algorithm for $3$-SAT such as DPLL \cite{davis-putnam, davis-logemann-loveland} (we have chosen the \textsc{MiniSat} implementation \cite{minisat}). The corresponding pseudo-code is shown in Algorithm \ref{alg:complete}.

\begin{algorithm}
	\begin{algorithmic}
	\Function{Complete-AdSAT}{$\phi_G$}  
		\ForAll{allowed $\mathcal{J}$} 
			\State run DPLL on $\phi_{\mathrm{G}}(\cdot, \mathcal{J})$
			\If{$\phi_{\mathrm{G}}(\cdot, \mathcal{J})$ is UNSAT}
				\State \Return UNSAT
			\EndIf
		\EndFor
		\State \Return SAT
	\EndFunction
	\end{algorithmic}
	\caption{A complete algorithm for AdSAT, implemented as a function returning ``SAT'' or ``UNSAT''. The loop is over the allowed configurations of negations in the sense that $N$ negations are fixed, following the argument in Sec.\ \ref{sec:preliminaries}.}
	\label{alg:complete}
\end{algorithm}

\begin{figure}
	\centering
	\includegraphics[width=0.48\columnwidth]{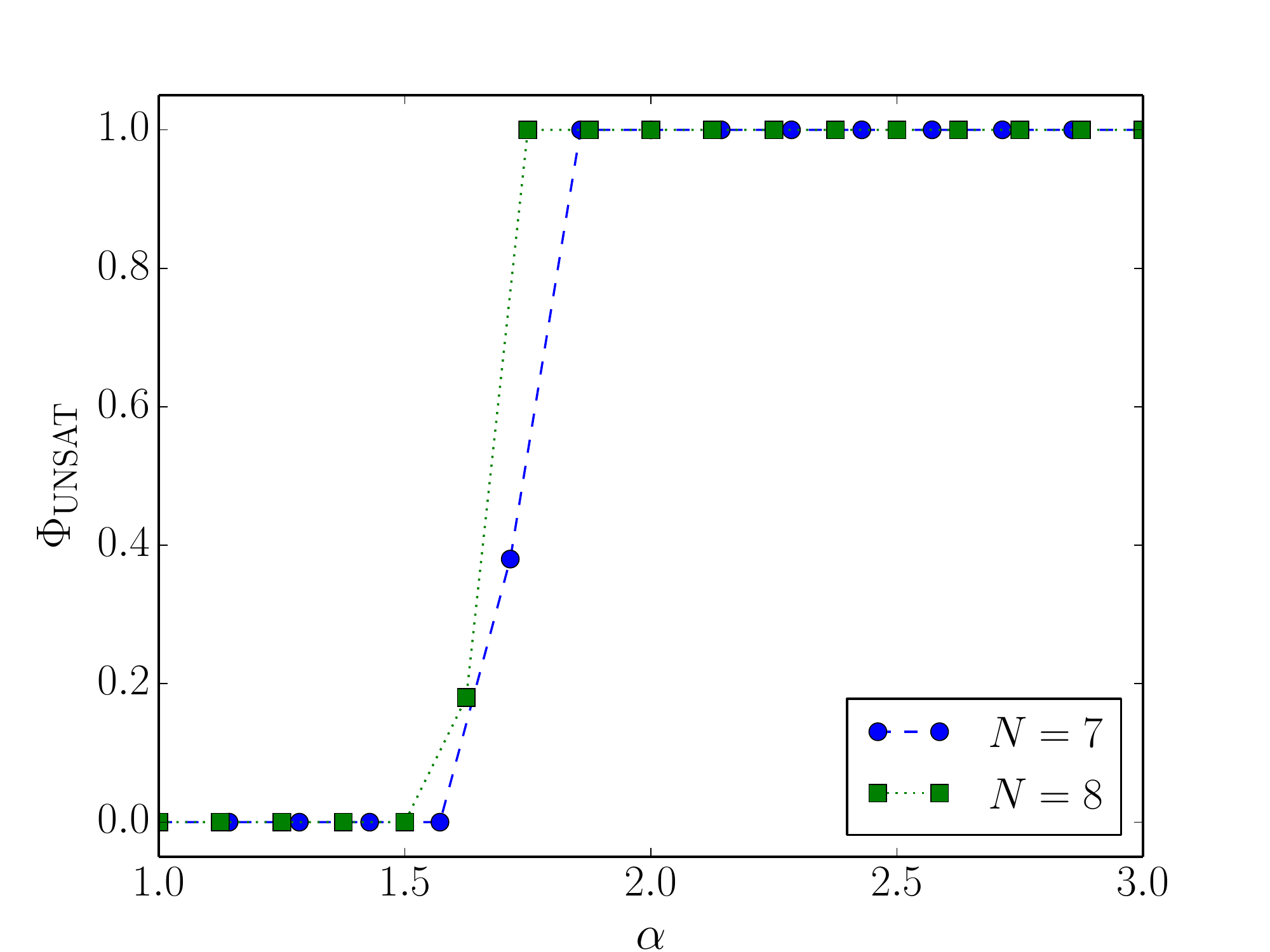}
	\caption{Fraction of UNSAT graphs $\Phi_{\mathrm{UNSAT}}$ with respect to $\alpha$ obtained with the complete algorithm. The total number of probed graphs for each value of $\alpha$ is $100$.}
	\label{fig:brute}
\end{figure}

In order to establish if the formula is SAT, the algorithm must try all the different $2^{K\alpha N}$ configurations of negations. Even taking into account the freedom to fix $N$ negations (see Sec.\ \ref{sec:preliminaries}), the effective number of configurations of negations is still $2^{(K\alpha-1)N}$. As a consequence, such an algorithm is not a viable option unless $N$ is very small. In Fig.\ \ref{fig:brute} we show the fraction of UNSAT graphs $\Phi_{\mathrm{UNSAT}}$ found with the complete algorithm vs $\alpha$, for $N=7$, $8$. Moving toward increasing values of $\alpha$, a transition seems to occur between a phase in which a graph is almost surely SAT, and a phase in which it is almost surely UNSAT. In fact, both for $N=7$ and for $N=8$ there is only a single value of $\alpha$ such that $\Phi_{\mathrm{UNSAT}}$ is not equal either to zero or to one. Even if one has to take into account that the spacing between two different viable values of $\alpha$ is of order $10^{-1}$ for such small values of $N$, it is also true that the transition window is expected to shrink for larger values of $N$. Overall, the presence of a sharp transition seems more likely than a smooth crossover between the two phases. However, given the smallness of $N$, nothing conclusive can be said about the value of $\alpha$ at which the transition occurs, as sizable corrections could be expected for larger values of $N$. 

%%%%%%%%%%%%%%%%%%%%%%%%%%%%%%%%%%%%%%%%%%%%%%%%%%%%%%%%%%%%%%%%%%%%%%%%%%%%%%%%%%%%%%%%%%%%%%%%%%%%

\section{Simulated annealing for AdSAT} 
\label{sec:annealing}
Given the impossibility to explore all the configuration space of negations one could think of moving in such a space along some preferred directions, until an UNSAT solution is found or a minimum of the number of SAT solutions is found. In particular one can introduce a cost function measuring the number of solutions and move in a direction in which the cost function decreases. 

Keeping in mind that the number of solutions is typically exponential in $N$ (for $\alpha<\alpha_c^{\mathrm{SAT}}$), a possible extensive cost function is the ``complexity'':
\begin{equation} \label{eq:compl}
	\Sigma_G(\mathcal{J}) = \frac{1}{N}\log_2(S_G(\mathcal{J}) + 1) \, ,
\end{equation}
where $S_G(\mathcal{J})$ is the number of solutions (we used \textsc{Relsat} \cite{relsat, relsat-web} to count the number of solutions) of the corresponding $3$-SAT problem with fixed negations $\mathcal{J}$ relative to the graph $G$. Note that $\Sigma_G(\mathcal{J}) = 0$ when the corresponding $3$-SAT problem is UNSAT. In this context, a move is simply the flip of a randomly selected negation. As noted in Sec.\ \ref{sec:preliminaries}, the number of effective negations is $3M - N$, meaning that the negations involving the first occurrences of each bit in the formula are fixed. 

In order to avoid getting stuck in a local minimum of the cost function one can use a Metropolis algorithm with a simulated annealing schedule. This is a stochastic process in which the move $\mathcal{J} \rightarrow \mathcal{J}'$ is always accepted if $\Sigma_G(\mathcal{J}') < \Sigma_G(\mathcal{J})$. On the other hand, if $\Sigma_G(\mathcal{J}') > \Sigma_G(\mathcal{J})$, the move can be still accepted if $e^{-\beta\left[\Sigma_G(	\mathcal{J}') - \Sigma_G(\mathcal{J})\right]} > \eta$, where $\eta$ is drawn uniformly between zero and one. Here $\beta$ plays the role of an inverse ``temperature''. In simulated annealing \cite{kirkpatrick-gelatt-vecchi}, the ``temperature'' progressively decreases. After testing some temperature schedules, we found that $\beta=2\sqrt{i}$ provides the best results, where $i=1, \ldots, I$ is the iteration number, and $I$ is the total number of iterations. We opt to start from a balanced configuration of negations, i.e.\ a configuration such that each bit is negated the same number of times, because these are much more constrained (they have typically much fewer solutions) than the typical random SAT formula. In Algorithm \ref{alg:annealing} (see Appendix \ref{sec:pseudo}) we show the pseudo-code for simulated annealing for AdSAT.

\begin{figure}
	\centering
	\includegraphics[width=0.48\columnwidth]{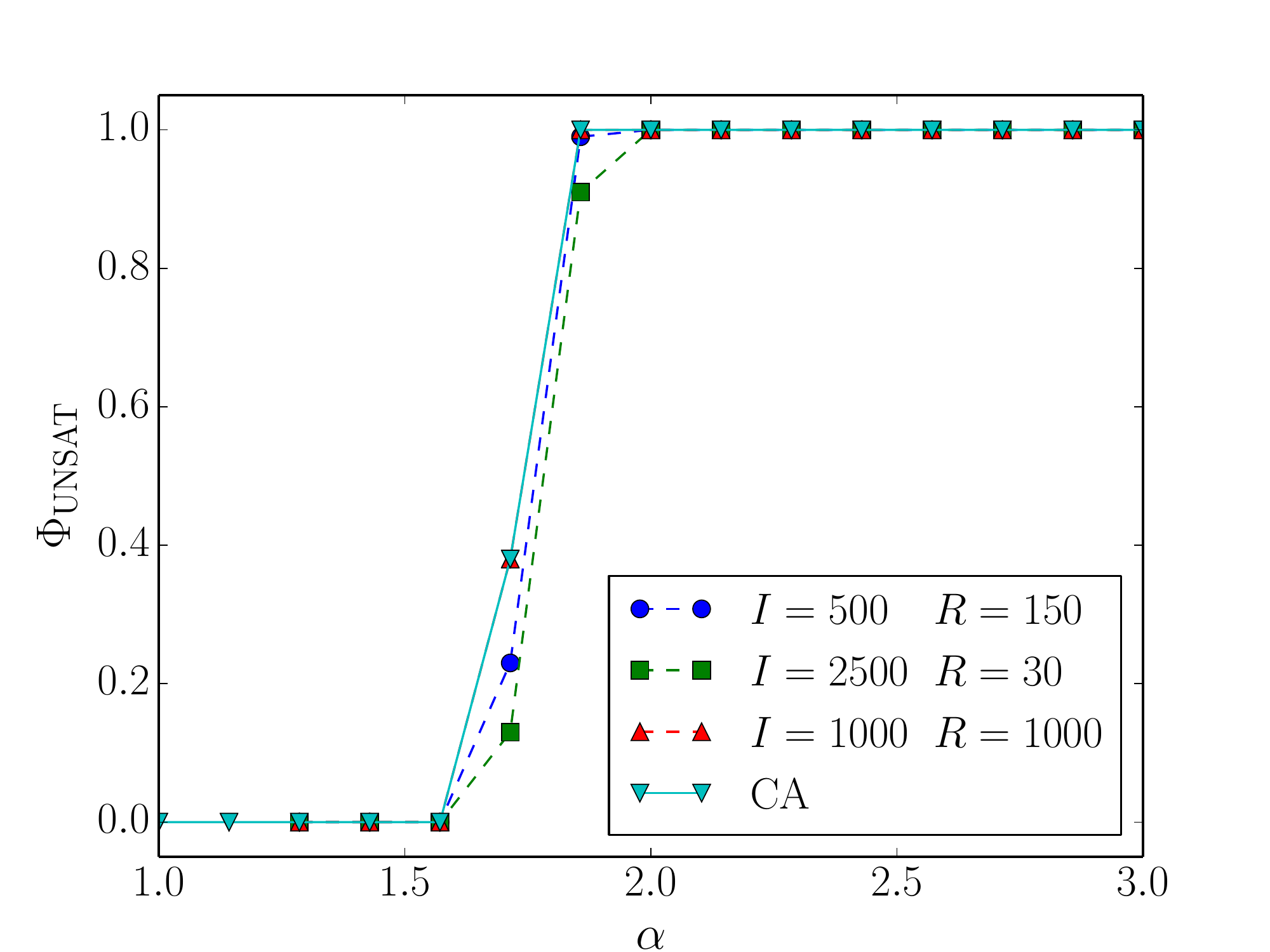}
	\includegraphics[width=0.48\columnwidth]{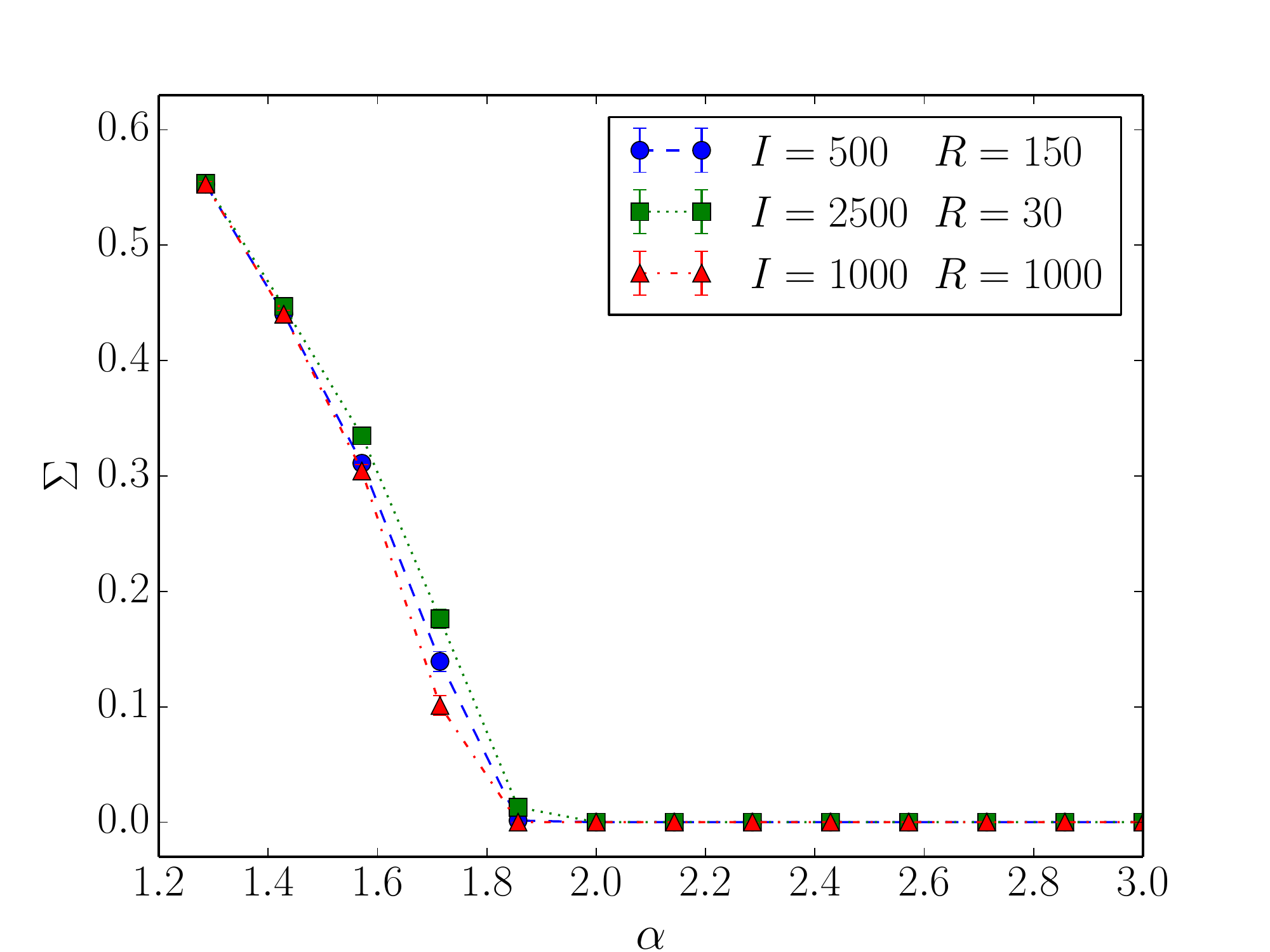}
	\includegraphics[width=0.48\columnwidth]{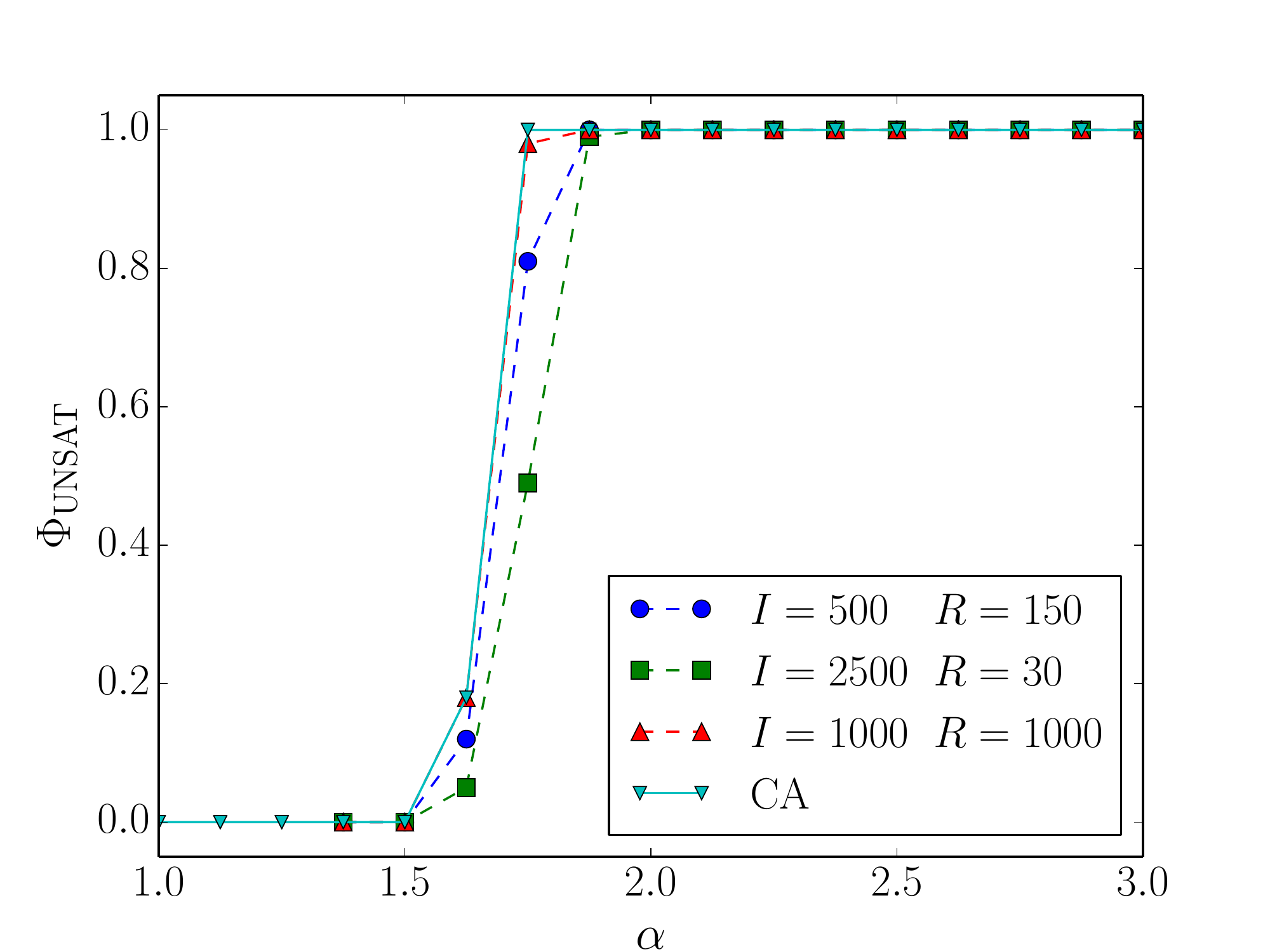}
	\includegraphics[width=0.48\columnwidth]{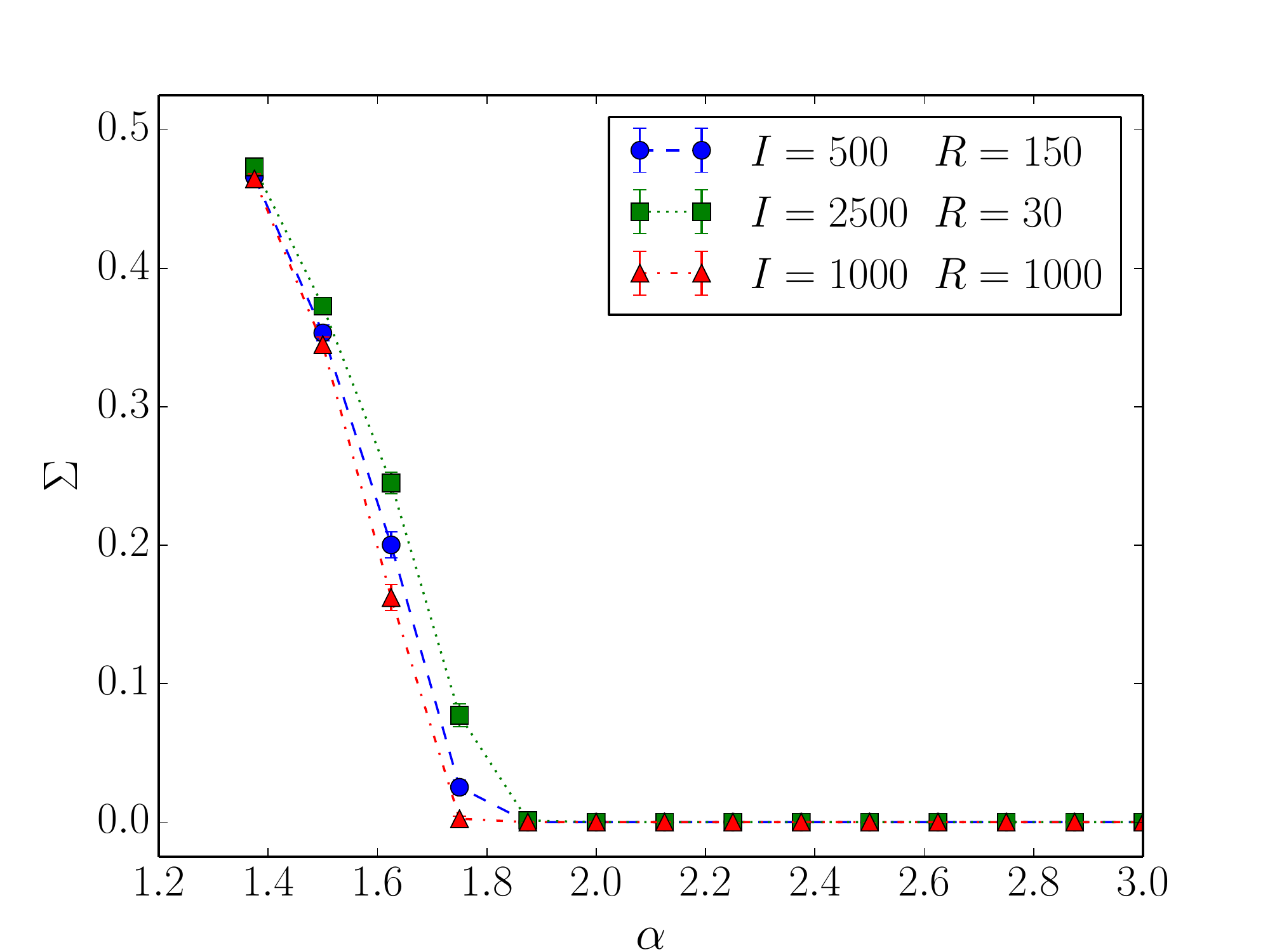}
	\caption{Comparison between the complete algorithm (CA) and simulated annealing for $N=7$ (top panels) and $N=8$ (bottom panels). Fraction of UNSAT graphs $\Phi_{\mathrm{UNSAT}}$ (left panels) and average minimal cost function $\Sigma$ (right panels) with respect to $\alpha$. Error bars are one standard deviation below and above the average, and within the symbol size. The total number of probed graphs for each value of $\alpha$ is $100$.}
	\label{fig:iter-reps}
\end{figure}

\begin{figure}
	\centering
	\includegraphics[width=0.48\columnwidth]{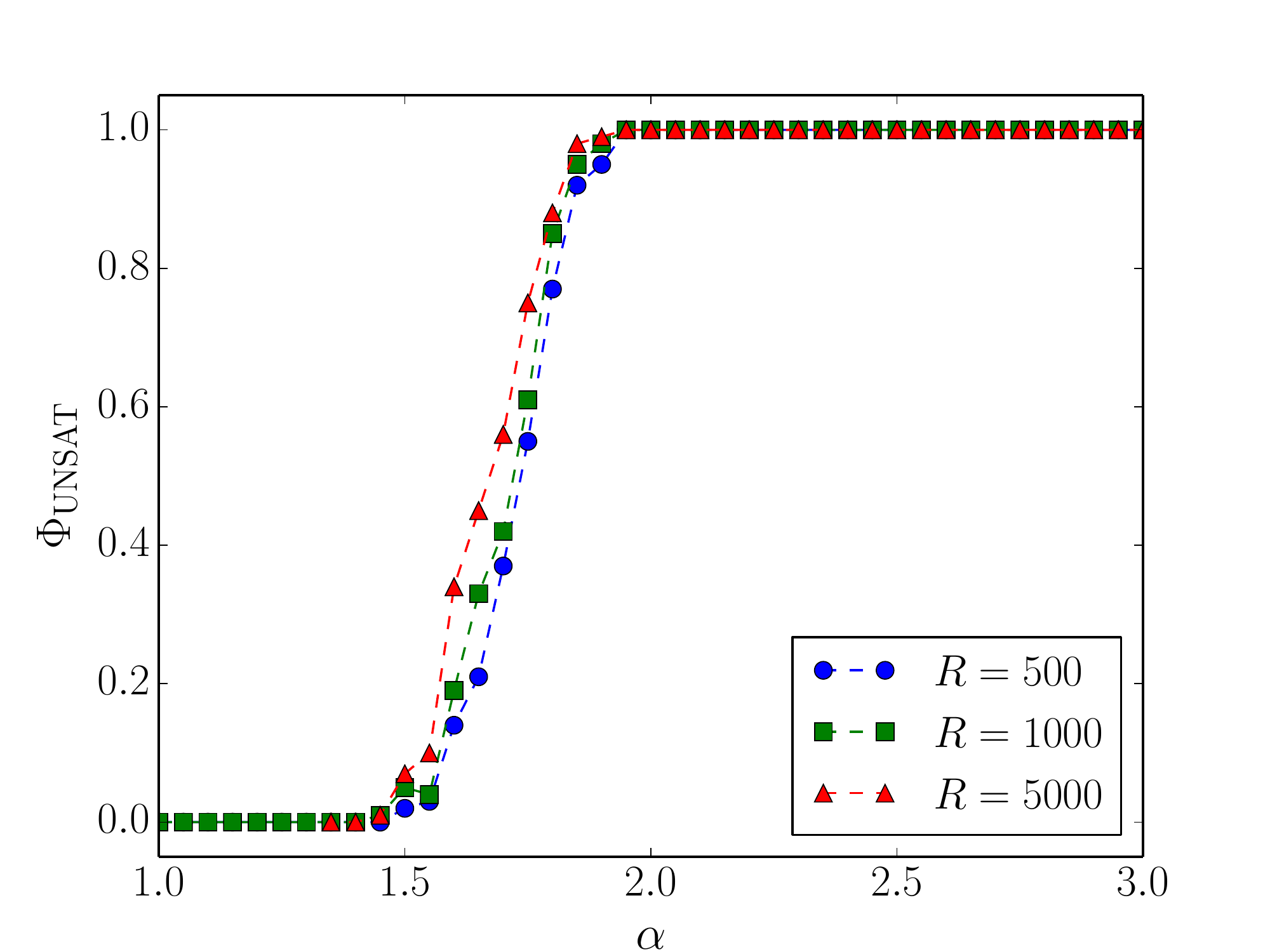}
	\includegraphics[width=0.48\columnwidth]{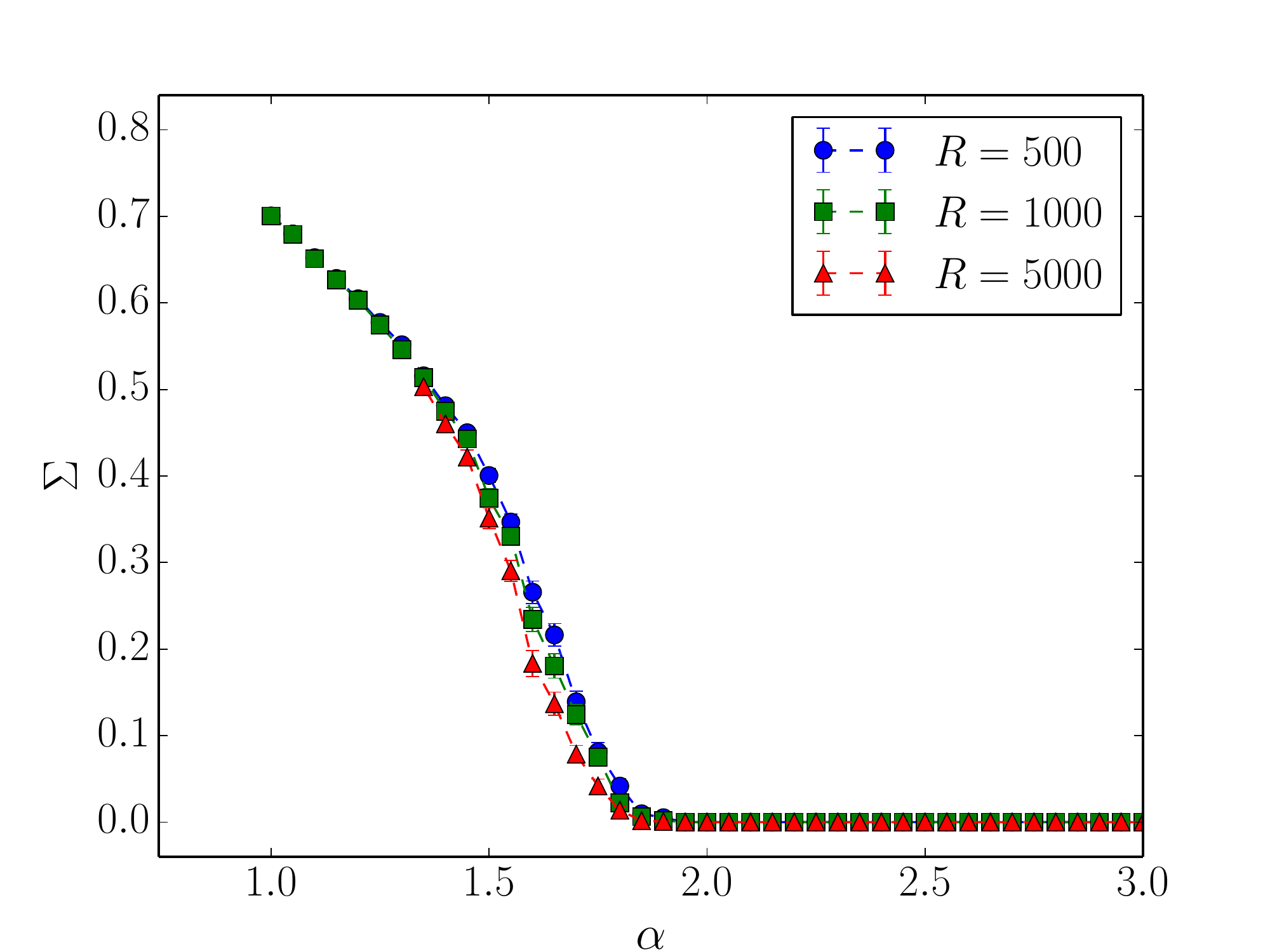}
	\caption{Fraction of UNSAT graphs $\Phi_{\mathrm{UNSAT}}$ (left panel) and average minimal cost function $\Sigma$ (right panel) with respect to $\alpha$ given by simulated annealing for $N=20$. Error bars are one standard deviation below and above the average, and within the symbol size. The total number of probed graphs for each value of $\alpha$ is $100$ and $I=500$.}
	\label{fig:n20-reps}
\end{figure}

A common extension of stochastic algorithms that is known to improve their efficiency is based on the concept of restarts. The basic idea is that the explored portion of the configuration space is not increased by increasing $I$, i.e.\ the length of a single path in such a space, but by increasing the number of paths, i.e.\ by restarting the algorithm from the beginning a certain number of times $R$. It can be na\"{i}vely expected that this extension is especially appropriate for simulated annealing, as a move implying $\Sigma_G(\mathcal{J}') > \Sigma_G(\mathcal{J})$ is less and less likely for lower and lower temperatures, and the chances to be stuck in a local minimum get larger and larger. In Algorithm \ref{alg:restarts} (see Appendix \ref{sec:pseudo}) we show the corresponding pseudo-code. In order to check for the relevance of introducing restarts we tested several combinations of $I$ and $R$ using both the fraction of UNSAT graphs $\Phi_{\mathrm{UNSAT}}$ and the minimal cost function averaged over the graphs:
\begin{equation}
	\Sigma = \overline{\min_{\mathcal{J}}\Sigma_G(\mathcal{J})} \, ,
\end{equation}
where the minimum is obviously over the configuration of negations explored by the algorithm. From Fig.\ \ref{fig:iter-reps} it is evident that keeping the total number of steps $IR$ constant, a better performance is achieved by increasing $R$ rather than $I$. The fact that it is necessary to use very large values of both $I$ and $R$ to obtain results comparable to those given by the complete algorithm, even for such small values of $N$, is a clear sign of the difficulty of the problem. This observation is confirmed by probing larger values of $N$. In fact, from Fig.\ \ref{fig:n20-reps} we see that for $N=20$ there is not a value of $R$ reachable under our computational limits such that either $\Phi_{\mathrm{UNSAT}}$ or $\Sigma$ start to saturate.

%%%%%%%%%%%%%%%%%%%%%%%%%%%%%%%%%%%%%%%%%%%%%%%%%%%%%%%%%%%%%%%%%%%%%%%%%%%%%%%%%%%%%%%%%%%%%%%%%%%%

\section{An improved stochastic algorithm} 
\label{sec:improved}
We now propose a more efficient way to explore the space of configurations for values of $\alpha$ close to the transition. Starting from a formula $\phi_G$ composed by $M$ clauses and a positive integer $\Delta M$ we build a sequence of formulas $\phi_1, \ldots, \phi_{\Delta M}$; defining for convenience $\phi_0 \equiv \phi_G$, $\phi_{s+1}$ is obtained simply by adding one (new) clause to $\phi_s$. For $\Delta M$ sufficiently large $\phi_{\Delta M}$ will be almost surely in the UNSAT phase, and will be easy to make it UNSAT using simulated annealing. Let $\mathcal{J}_{\Delta M}$ be the configuration of negations that makes $\phi_{\Delta M}$ UNSAT. Assuming that the configuration of negations that makes $\phi_{\Delta M - 1}$ UNSAT (or for which its number of solutions of is minimal) is not ``far'' from $\mathcal{J}_{\Delta M}$, we try to make $\phi_{\Delta M - 1}$ UNSAT using simulated annealing and starting from the configuration of negation $\mathcal{J}_{\Delta M}$, but with the last clause removed. Clearly the goodness of such approach has to be evaluated a posteriori, and in particular by comparing it with the simulated annealing introduced in Sec.\ \ref{sec:annealing}. This procedure is recursively iterated proceeding from $\phi_{\Delta M}$ to $\phi_0$ and, while passing from $\phi_{s+1}$ to $\phi_s$, keeping the configuration of negations that has made $\phi_{s+1}$ UNSAT or, if $\phi_{s+1}$ has not been made UNSAT, the last accepted configuration in the annealing of $\phi_{s+1}$ (which might not necessarily be the configuration with the minimum number of solutions for $\phi_{s+1}$). Once at $\phi_0$ we run the simulated annealing normally to minimize the number of solutions. In such a way the full annealing is divided in $\Delta M + 1$ chunks each of (maximum) $I$ steps, for a total of $(\Delta M + 1) I$ steps; for each chunk, if $\phi_s$ is made UNSAT during the annealing, the effective number of steps will be lower. For consistency we use alternately balanced formulas, i.e.\ in which each bit is alternately negated and not negated, so that the balance holds even after removing the clauses added to $\phi_0$. A subtle point is how to handle the temperature schedule of the annealing while passing from $\phi_{s+1}$ to $\phi_s$. The schedule used has the same endpoints of the schedule described in Sec.\ \ref{sec:annealing}, in the sense that the first chunk starts from $\beta = 0$ and the last chunk ends at $\beta = 2 \sqrt{I}$. We explicitly note that in the final chunk (the annealing starting from $\phi_0$) the temperature is lower than it would be if the final chunk were the only chunk (as in Sec.\ \ref{sec:annealing}). The corresponding pseudo-code is shown in Algorithm \ref{alg:improved} (see Appendix \ref{sec:pseudo}).

\begin{figure}
	\centering
	\includegraphics[width=0.48\columnwidth]{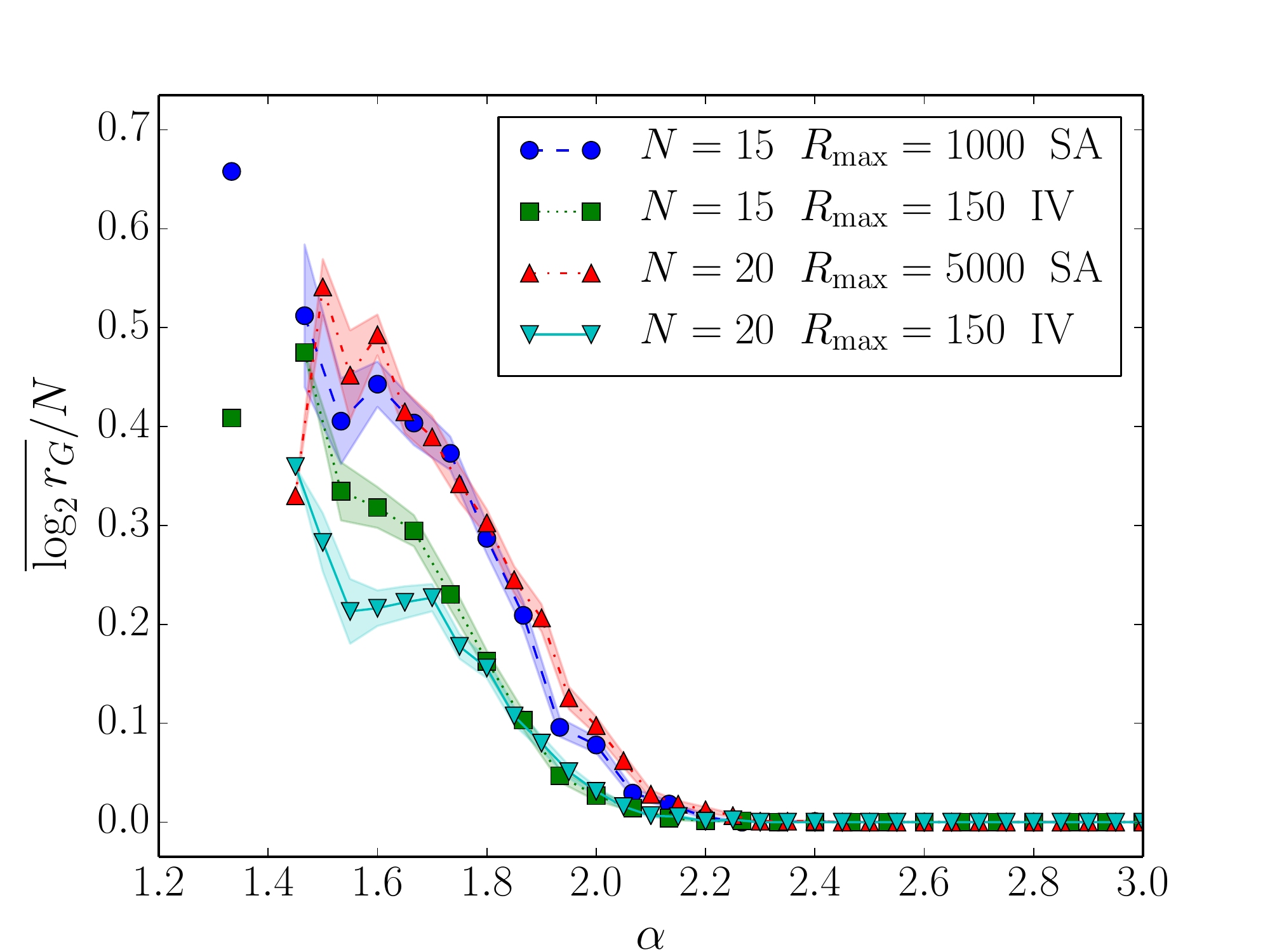}
	\caption{Comparison between simulated annealing (SA) and its improved variant (IV) for $N=15$ and $N=20$. Averaged logarithm of the number of restarts after which a graph is made UNSAT $\overline{\log_2 r_G} /N$ with respect to $\alpha$. Semi-transparent regions are one standard deviation below and above the average. The total number of probed graphs for each value of $\alpha$ is $100$ and $I=500$.}
	\label{fig:alg-comp}
\end{figure}

In order to compare the performance of the simple simulated annealing with that of its improved variant we look precisely at the value of the restart $r_G$ after which the formula has been made UNSAT; clearly $r_G \leq R$. In Fig.\ \ref{fig:alg-comp} we plot $\overline{\log_2 r_G} /N$ (the over line stands for the average over the graphs that have been made UNSAT) vs $\alpha$. It is clear that the improved algorithm is able to provide an equal or better performance than simulated annealing, and such a gain is more and more noticeable passing from $N=15$ to $N=20$, especially for $\alpha < 1.8$. 

Subsequently, we analyze how the critical value of $\alpha$ changes with respect to the number of restarts in the improved algorithm. It is convenient to define
\begin{equation}
	\gamma = N/\log_2 R \, ,
\end{equation}
which contains the information on the number of restarts in an appropriate way. In fact, one expects the complexity of the problem, indicated by the value of $R$ necessary to obtain convergence, to scale exponentially with $N$, in analogy with what happens with the \textsc{WalkSAT} and other local algorithms for the $K$-SAT problem \cite{papadimitriou, schoening-1, schoening-2}. Therefore one should observe a critical value of $\gamma$ below which the results do not depend on $\gamma$. 

\begin{figure}
	\centering
	\includegraphics[width=0.48\columnwidth]{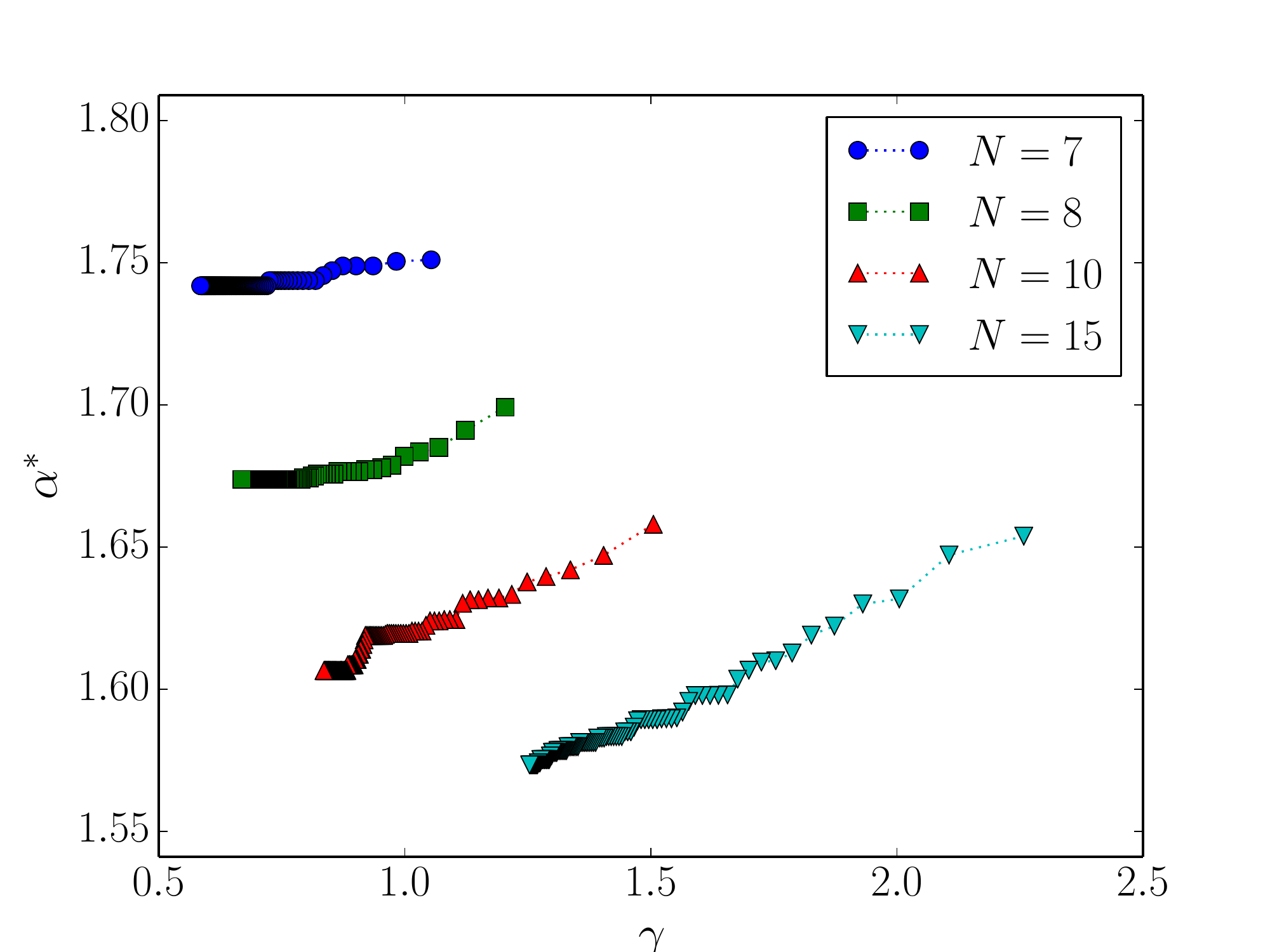}
	\caption{Interpolated critical value of the density of clauses $\alpha^*_N$ with respect to $\gamma$ given by the improved algorithm for $N=7$, $8$, $10$, $15$. The total number of probed graphs for each value of $\alpha$ is $100$ and $I=2000$. $R_{\mathrm{max}}$ is the number of restarts corresponding to the left endpoint of the curves, and is equal to $4000$ for all the curves.}
	\label{fig:n-comp}
\end{figure}

The result that we are most interested in is the critical value $\alpha^*_N$ for which the fraction of SAT and UNSAT formulas is the same ($\Phi_{\mathrm{UNSAT}}=\Phi_{\mathrm{SAT}} = 1/2$).\footnote{In practice it is determined in the following way: we compute $|\Phi_{\mathrm{UNSAT}} - 1/2|$ for all the values of $\alpha$, rank them in ascending order and select the first five values; then we make a linear interpolation between these points and use the interpolating function to compute $\alpha^*$, the value of $\alpha$ corresponding to $\Phi_{\mathrm{UNSAT}} = 1/2$.} We expect that as $R$ is increased for a fixed $N$, the curve $\alpha^*_N(\gamma)$ should bend and a plateau will intercept the $\alpha^*_N$ axis at the predicted ``critical" $\alpha$ for the given $N$, as it is evident from the curves for $N=7$, $8$ and $10$ in Fig.\ \ref{fig:n-comp}. The plateau therefore signals the reaching of the asymptote and the sufficiency of the number of restarts $R$. Actually for both $N=7$ and $8$ at the plateau one completely recovers the results given by the complete algorithm. Increasing $N$ the plateau shifts towards smaller and smaller values of $\alpha^*_N$. As a consequence, all the curves at finite $N$ would need a negative shift in $\alpha^*_N$ to collapse on the theoretical limit curve for $N$ going to infinity, testifying that the thermodynamic limit of $\alpha^*$ is at smaller values than those observed here. Using the values of $I$ and $R$ practically accessible to us, the curve for $N=15$ does not reach a plateau, but it already extends up to values of $\alpha^*_{15}$ well below the plateau of the curve for $N=10$.  

Actually, while the position of the plateaux depends only on $N$, the position of the knee depends also, although weakly, on the number of iterations $I$ per restart. In fact for larger values of $I$ the knee should move towards larger values of $\gamma$. 

As regards $N=20$, due to our present computational limits, we focus on the single value $\alpha = 1.6$, which is right above the transition for $N=15$ (with $64$ UNSAT graphs). With $I=2000$ and $R=18000$ we are able to make UNSAT only $62$ graphs for $N=20$. One could be tempted to consider $I=2000$ not sufficiently large for $N=20$, but increasing $I$ while keeping $R$ large enough is not a viable option due to our current computational limits. Such a result could depend either on the fact that the number of needed restart is expected to scale exponentially in the critical region, or to an inversion point, in the sense that the threshold is starting to move towards larger values of $\alpha$.

Since disentangling the two effects is beyond our reach at the moment, we try to gain more information by investigating graphs of larger size, but restricting to values of $\alpha$ close to $3$. In all the investigated cases we are able to make UNSAT $100$ graphs out of $100$. In particular, for $\alpha=3$ and $N=150$ all graphs are UNSAT using $I=2000$ after a maximum of $45$ restarts, for $\alpha=2.75$ and $N=100$ using $I=2000$ after a maximum of $194$ restarts, and for $\alpha=2.70$ and $N=100$ using $I=4000$ after a maximum of $194$ restarts. Looking more closely, passing from $\alpha=2.75$ to $\alpha=2.70$ there is a sizeable drop in the number of graphs made UNSAT for $I=1000$ (respectively $84$ and $47$), which could be a hint of the beginning of the critical region. Anyway, since for larger values of $I$ there is not much difference (for $\alpha=2.70$ already $97$ graphs are made UNSAT using $I=2000$) it is more likely that lower values of $I$ are not sufficiently large when we enter the region in which the graphs become ``tougher''.
The small number of restarts is quite indicative of the fact that $\alpha=2.70$ is still quite above the critical value (in the critical region, the number of restarts needed to decide the formula is exponential in $N$) \footnote{Being optimistic, one could conjecture that the complexity of the problem is $\gamma\sim 1$ as seen for the low $N$ numerics. This means that, assuming a sufficient number of iterations (probably in the millions), the number of restarts necessary to get a converged $\alpha_N^*$ for $N=150$ is about $2^{150}\sim 10^{45}$.}.
As a consequence, a value of $\alpha_c > 2$ in the thermodynamic limit is conceivable only admitting the possibility of a very large $1/N$ correction (logarithmic or a small power). If the leading corrections are $1/N$ as usual in mean-field glasses, to reconcile with the theoretical expectations of $\alpha_c=3.39$ \cite{castellana-zdeborova}, the coefficient should be $\simeq 200$, which has to be compared with one as $\alpha$ is dimensionless. Moreover, an even larger subleading ($1/N^2$) coefficient with a negative sign should be present, since we observe a decreasing of $\alpha^*_N$ with $N$ for all the accessible values of $N$. At the moment, we do not know the possible reasons for the discrepancy with the result $\alpha^*=3.39$ in \cite{castellana-zdeborova}. 

The numerics performed in \cite{castellana-zdeborova} (simulated annealing without restarts) for the ensemble of random regular graphs (an $L$-regular graph is a graph in which every variable belongs exactly to $L$ clauses, so that $K M = L N$) seems to suggest a smaller value of the transition threshold $\alpha_c \simeq 2.33$ ($L=7$), but with finite-size corrections pointing towards larger values. In particular, $\Phi_{\mathrm{UNSAT}} = 1$ for $L>7$, while $\Phi_{\mathrm{UNSAT}} \simeq 0.1$ for $L=7$ and $N=54$. For $L=6$ ($\alpha = 2$) $\Phi_{\mathrm{UNSAT}} \simeq 0.5$ for $N=18$, $\Phi_{\mathrm{UNSAT}} \simeq 0.15$ for $N=27$, and $\Phi_{\mathrm{UNSAT}} = 0$ for larger values of $N$. $\Phi_{\mathrm{UNSAT}} = 0$ for smaller values of $L$. Our algorithm is able to outperform the numerics in \cite{castellana-zdeborova} finding $\Phi_{\mathrm{UNSAT}} = 0.58$ for $L=7$ and $N=54$. For $L=6$ we find $\Phi_{\mathrm{UNSAT}} = 0.2$ for $N=36$, $\Phi_{\mathrm{UNSAT}} = 0.94$ for $N=27$, and $\Phi_{\mathrm{UNSAT}} = 1$ for $N=18$. Consequently, also in this case it is impossible to distinguish if this effect is due to a genuine shift of $\alpha_c$, or to the increasing difficulty to make larger and larger graphs UNSAT. In \cite{castellana-zdeborova} the authors explain this discrepancy (the analytical value $\alpha_c$ is around 3.66 for regular graphs) with the presence of strong pre-asymptotic effects. 

Using our improved algorithm to probe regular graphs we see a transition threshold shifting towards smaller values of $\alpha$ for increasing $N$, again in contrast with \cite{castellana-zdeborova}. In particular, using $I=2000$ and $R=4000$, for $N=9$ the curve $\alpha^*_N(\gamma)$ reaches the plateau around $1.81$, while for $N=12$ it arrives to $1.79$ without reaching the plateau. Let us point out that determining the value of $\alpha^*_N$ for regular graphs is certainly more problematic, due to the large granularity in the values of $\alpha$. In fact the allowed values of $\alpha$ are $1, 4/3, 5/3, 2, 7/3, \ldots$ if $N$ is a multiple of $3$, and only the integers otherwise. 

Finally, we would like to comment on the configuration of negations that minimize $\Sigma$. As was also pointed out in \cite{castellana-zdeborova}, it is reasonable to expect that such a configuration is balanced in the thermodynamic limit. However, even after requesting that a configuration is balanced, there is still a lot of freedom to set the negations. There is plenty of room for more complex, long-range correlations between negations to play a role. In other words, one expects the balancing to be a necessary condition for a configuration minimizing $\Sigma$, but not a sufficient one, giving then only an \emph{upper bound} to the AdSAT threshold, as we find indeed in our numerics. Looking at the configurations obtained at the end of our improved algorithm for small values of $N$ (i.e.\ far from the thermodynamic limit) we actually find that, when the graph becomes UNSAT only after a nontrivial number of iterations, the fraction of balanced configurations could even be very small. For $N = 100, 125, 150$ and $\alpha = 3$ we find a larger fraction of balanced configurations (around $75\%$) and slightly increasing with $N$. Therefore, at small values of $N$ (meaning $N\lesssim 200$) these other correlations play even a larger role than balancing, leading to best solutions that are \emph{unbalanced}. We think that this effect will go away in the thermodynamic limit, but this is a clear indication that other conditions are playing a prominent role making the restriction to balanced instances necessary but not sufficient. This makes the value of $\alpha_c$ found in \cite{castellana-zdeborova} an upper bound rather than the exact value of the AdSAT transition threshold.

%%%%%%%%%%%%%%%%%%%%%%%%%%%%%%%%%%%%%%%%%%%%%%%%%%%%%%%%%%%%%%%%%%%%%%%%%%%%%%%%%%%%%%%%%%%%%%%%%%%%

\section{Conclusions} 
\label{sec:conclusions}

We have numerically investigated AdSAT, a quantified boolean formula problem which originated from the study of the quantum analogue of SAT. Previous investigations in \cite{castellana-zdeborova} have proposed a threshold at $\alpha_c=3.39$. On one hand numerics on small system sizes (up to $N=15$) suggest $\alpha_c<1.6$. On the other hand, we are able to exclude the region $\alpha_c > 3$ ($\alpha_c > 2.70$) for systems of size up to $N=150$ ($N=100$) upon the assumption of regular $1/N$ corrections to the critical values of $\alpha^*_N$. As already pointed out, even considering the present limits on the numerically accessible values of $N$, these results could be compatible with the analytical results in \cite{castellana-zdeborova} only if unusually large finite-size corrections were present.

Moreover, our upper bound on $\alpha_c$ for AdSAT is also much closer to the numerical upper bound found in \cite{laumann2010random} for the threshold of 3-QSAT, signaling that the difference between the quantum problem and the classical one might not be so striking.

Improving on the algorithms would be much desirable as AdSAT is a member of the quantified boolean problems family, which is only recently starting to get the attention it deserves. One possibility would be to use a belief propagation supported heuristic algorithm as devised in \cite{zhang2012message} to solve AdSAT. We hope to do this in the future, together with a deep analysis of the nature of the configuration of negations minimizing $\Sigma$, which, as highlighted at the end of Sec.\ \ref{sec:improved}, is certainly a direction worth pursuing.

%%%%%%%%%%%%%%%%%%%%%%%%%%%%%%%%%%%%%%%%%%%%%%%%%%%%%%%%%%%%%%%%%%%%%%%%%%%%%%%%%%%%%%%%%%%%%%%%%%%%

\begin{acknowledgments}
The authors would like to thank Lenka Zdeborov\'{a} for many discussions in the early stages of this work. A.S.\ also thanks Roderich Moessner for many useful discussions.
D.N.\ was supported by Frank Verstraete's ERC grant QUERG and thanks the Slovak Research and Development Agency grant APVV-0646-10 COQI.
M.B.\ heartily thanks Maria Valentina Carlucci for her dedicated support.
\end{acknowledgments}

%%%%%%%%%%%%%%%%%%%%%%%%%%%%%%%%%%%%%%%%%%%%%%%%%%%%%%%%%%%%%%%%%%%%%%%%%%%%%%%%%%%%%%%%%%%%%%%%%%%%

\appendix

\section{2-AdSAT is in P} 
\label{sec:adsat2}

We now show that adversarial 2-SAT (2-AdSAT) is in $\mathbf{P}$, i.e.\ we prove that we can efficiently find out whether for a given graph $G$ of clauses (each involving 2 bits), an adversary can choose the negations so that the formula $\phi_G(x,\mathcal{J})$ (\ref{eq:formula}) is UNSAT for any choice of $x$.

Note that any disconnected subgraphs can be analyzed by themselves and that any tree-like (no loops) part of the graph can be cut off, as any 2-SAT instance on a tree is satisfiable for any value of the bit at the root of the tree.  

\begin{figure}
	\centering
	\includegraphics[width=8.5cm]{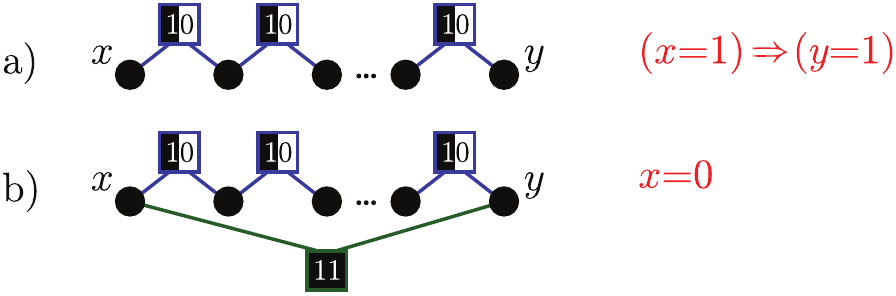}
	\caption{For 2-AdSAT, the adversary can choose the negations so that he a) creates an implication, b) fixes a bit.}
	\label{fig:adsat2gadgets}
\end{figure}

The adversary now has two simple ways to set up the negations. First, he can use an ``implication line'' (see Fig.\ \ref{fig:adsat2gadgets}a) which works as $(x=1) \Rightarrow (y=1)$. It is a line of clauses with the negations set so that a clause involving two successive bits $b_i$ and $b_{i+1}$ on the line is $(b_i\oplus 1)\vee b_{i+1}$. When all of these clauses are true, $x=1$ implies $y=1$. Note that if we choose $(b_i\oplus 1)\vee (y\oplus 1)$ in the last clause, we create the implication line $(x=1) \Rightarrow (y=0)$. The adversary can similarly set the implication lines $(x=0) \Rightarrow (y=0)$ and $(x=0) \Rightarrow (y=1)$.

Second, the adversary can use a ``bit-fixing cycle'' (see Fig.\ \ref{fig:adsat2gadgets}b) which sets $x=0$. It is made from an implication line $(x=1)\Rightarrow (y=1)$ closed by the final clause $\overline{x}\vee \overline{y}$. These can be simultaneously true only for $x=0$. Analogously, the adversary can choose to fix $x=0$.

\begin{figure}
	\centering
	\includegraphics[width=\columnwidth]{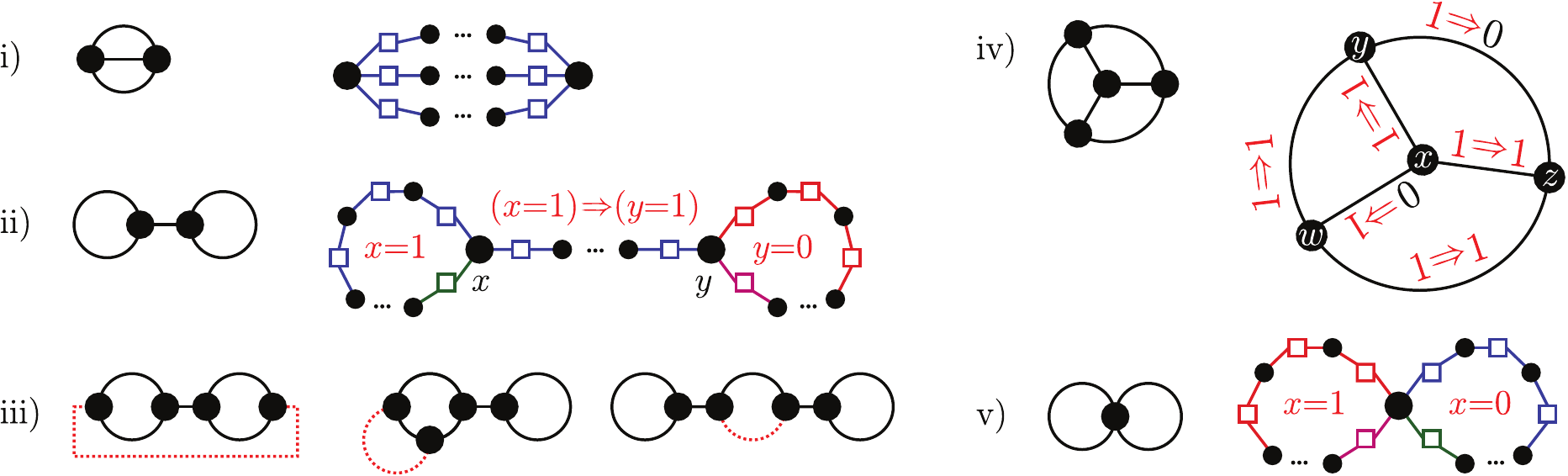}
	\caption{Solving 2-AdSAT divides into a few easily checkable cases (assuming a connected component with pruned tree-like parts). Only the graph of the type i) is adversarially satisfiable (SAT for any choice of negations), the others are UNSAT.}
	\label{fig:adsat2}
\end{figure}

We can now apply these tools to the remainder of the graph after dividing it into disconnected components and pruning the tree-like parts. If there is only a single loop (no node has degree more than 2), the instance is trivially SAT, no matter how hard the adversary tries. Let us then look at the five remaining interesting cases that can happen within a connected component.

\begin{enumerate}
	\item The largest node degree is 3. If there are exactly two nodes with degree 3, connected via 3 paths as in Fig. \ref{fig:adsat2}i), the adversarial 2-SAT instance is SAT. One can rule out most of the possible solutions by setting up implication lines on the paths. However, this allows us to set only three implication lines for the 2 bits (nodes with degree 3), which is not enough to rule all 4 possible bit assignments of the two degree-3 nodes. Such an adversarial 2-SAT instance is thus SAT.
	\item The largest node degree is 3. There are exactly two nodes with degree 3, connected via a single path as in Fig.\ \ref{fig:adsat2}ii). The adversarial 2-SAT instance is now UNSAT, as we can fix the bits to $x=1$, $y=0$, and connect them via an implication line $(x=1)\Rightarrow (y=1)$ that rules out this possibility. 
	\item The largest node degree is 3. There are exactly 4 nodes with degree 3, and the graph contains the case ii) as a subgraph (see Fig.\ \ref{fig:adsat2}iii). This can be tested by removing lines (the red dotted lines in Fig.~\ref{fig:adsat2}iii) and seeing what we get. The adversary then deals with the subgraph described in ii), making it UNSAT.
	\item The largest node degree is 3. There are exactly 4 nodes of degree 3, and when we remove any of the lines, we recover the case i). This means our graph is depicted in Fig.\ \ref{fig:adsat2}iv). We can now set up three implication lines
\begin{eqnarray*} 
	(x=1) &\Rightarrow& (y=1) \, , \\
	(x=1) &\Rightarrow& (z=1) \, , \\
	(y=1) &\Rightarrow& (z=0) \, . 
\end{eqnarray*}
This fixes $x=0$, because $x=1$ implies $y=z=1$, which is inconsistent with the last implication. We now set up three more implication lines
\begin{eqnarray*} 
	(x=0) &\Rightarrow& (w=1) \, , \\
	(w=1) &\Rightarrow& (y=1) \, , \\
	(w=1) &\Rightarrow& (z=1) \, . 
\end{eqnarray*}
Assuming now that $x=0$, we have $w=y=z=1$. However, because we already know that $y=z=1$ is inconsistent with one of the implications above, such an adversarial 2-SAT instance is UNSAT. Note that starting with more than 4 nodes with degree-3 in a connected component, the 2-AdSAT instance is UNSAT, as it maps either to iii) or iv) by removing lines until we end up with a connected component with exactly four degree-3 nodes. 
	\item Finally, we are left with the cases which contain a node with degree at least 4. The graph then contains a subgraph depicted in Fig.\ \ref{fig:adsat2}v). The adversary uses the two loops to fix the bit $x$ to be 0 and 1, making the instance UNSAT.
\end{enumerate}

%%%%%%%%%%%%%%%%%%%%%%%%%%%%%%%%%%%%%%%%%%%%%%%%%%%%%%%%%%%%%%%%%%%%%%%%%%%%%%%%%%%%%%%%%%%%%%%%%%%%

%\clearpage
\section{Pseudo-codes}
\label{sec:pseudo}
Here we list the pseudo-codes for simulated annealing for AdSAT (Algorithm\ \ref{alg:annealing}), its extension with restarts (Algorithm\ \ref{alg:restarts}), both discussed in Sec.\ \ref{sec:annealing}, and the improved variant (Algorithm\ \ref{alg:improved}), discussed in Sec.\ \ref{sec:improved}.

\begin{algorithm}
	\begin{algorithmic}
	\Function{SA-AdSAT}{$\phi_G, \mathcal{J}, I$}
		\State $\Sigma_G(\mathcal{J}) \gets$ complexity of $\phi_G(\cdot, \mathcal{J})$
		\State $\Sigma_G^* \gets \Sigma_G(\mathcal{J})$
		\If{$\Sigma_G(\mathcal{J}) = 0$}
			\State \Return 0, $\mathcal{J}$
		\EndIf
		\For{$t=1,\ldots,I$}
			\State $\beta \gets 2 \sqrt{t}$
			\State obtain $\mathcal{J}'$ flipping a random (allowed) negation in $\mathcal{J}$ 
			\State $\Sigma_G(\mathcal{J}') \gets$ complexity of $\phi_G(\cdot, \mathcal{J}')$
			\If{$\Sigma_G(\mathcal{J}') = 0$}
				\State \Return 0, $\mathcal{J}'$
			\ElsIf{$0 < \Sigma_G(\mathcal{J}') < \Sigma_G(\mathcal{J})$}
				\State $\mathcal{J} \gets \mathcal{J}'$ 
				\State $\Sigma_G(\mathcal{J}) \gets \Sigma_G(\mathcal{J}')$
				\If{$\Sigma_G(\mathcal{J}') < \Sigma_G^*$}
					\State $\Sigma_G^* \gets \Sigma_G(\mathcal{J}')$
				\EndIf
			\Else
				\State $\eta \gets$ random number in $[0, 1]$
				\If{$e^{-\beta \left[\Sigma_G(\mathcal{J}') - \Sigma_G(\mathcal{J}) \right]} > \eta$}
					\State $\mathcal{J} \gets \mathcal{J}'$ 	
					\State $\Sigma_G(\mathcal{J}) \gets \Sigma_G(\mathcal{J}')$
				\EndIf
			\EndIf
		\EndFor
		\State \Return $\Sigma_G^*$, $\mathcal{J}$
	\EndFunction
	\end{algorithmic}
	\caption{Simulated annealing for AdSAT, implemented as a function returning both the minimum value of the complexity $\Sigma_G^*$ found and the last accepted configuration of negations $\mathcal{J}$.}
	\label{alg:annealing}
\end{algorithm}

\begin{algorithm}
	\begin{algorithmic}
	\Function{Restart-SA-AdSAT}{$\phi_G, R, I$}
		\For{$r=1,\ldots,R$}
			\State pick a random balanced config. of negations $\mathcal{J}$
			\State $\Sigma_G, \, \tilde{\mathcal{J}} \gets$ \textsc{SA-AdSAT}($\phi_G, \mathcal{J}, I$) 
			\If{$\Sigma_G = 0$}
				\State \Return 0
			\EndIf
			\If{$r = 1$}
				\State $\Sigma_G^* \gets \Sigma_G$
			\ElsIf{$r >1$ and  $\Sigma_G < \Sigma_G^*$}
				\State $\Sigma_G^* \gets \Sigma_G$
			\EndIf 
		\EndFor
		\State \Return $\Sigma_G^*$	
	\EndFunction
	\end{algorithmic}
	\caption{Simulated annealing for AdSAT with restarts, implemented as a function returning  the minimum value of the complexity $\Sigma_G^*$ found.}
	\label{alg:restarts}
\end{algorithm}

\begin{algorithm}
	\begin{algorithmic}
	\Function{IV-AdSAT}{$\phi_G, \Delta M, R$}
		\State $M \gets$ number of clauses in $\phi_G$
		\For{$r=1,\ldots,R$}
			\State $s \gets M + \Delta M$
			\State $\phi_{s} \gets \phi_G$ expanded with $\Delta M$ additional clauses
			\State pick a random alternately balanced configuration of negations $\mathcal{J}_s$
			\While{$s > 0$}
				\State $\tilde{\Sigma}, \, \mathcal{J}_s \gets$ \textsc{SA-AdSAT}($\phi_s$, $\mathcal{J}_s, I$) 
				\State $\phi_{s-1} \gets$ $\phi_s$ with the last clause removed
				\State $\mathcal{J}_{s-1} \gets \mathcal{J}_s$ with the last clause removed
				\State $s \gets s-1$
			\EndWhile
			\State $\Sigma_G, \, \tilde{\mathcal{J}} \gets$ \textsc{SA-AdSAT}($\phi_0$, $\mathcal{J}_0, I$) 
			\If{$\Sigma_G = 0$}
				\State \Return 0
			\EndIf
			\If{$r = 1$}
				\State $\Sigma_G^* \gets \Sigma_G$
			\ElsIf{$r >1$ and  $\Sigma_G < \Sigma_G^*$}
				\State $\Sigma_G^* \gets \Sigma_G$
			\EndIf 
		\EndFor
		\State \Return $\Sigma_G^*$	
	\EndFunction
	\end{algorithmic}
	\caption{Improved variant of the simulated annealing for AdSAT with restarts, implemented as a function returning  the minimum value of the complexity $\Sigma_G^*$ found.}
	\label{alg:improved}
\end{algorithm}

%%%%%%%%%%%%%%%%%%%%%%%%%%%%%%%%%%%%%%%%%%%%%%%%%%%%%%%%%%%%%%%%%%%%%%%%%%%%%%%%%%%%%%%%%%%%%%%%%%%%

\bibliography{adsat}

\begin{thebibliography}{29}%
\makeatletter
\providecommand \@ifxundefined [1]{%
 \@ifx{#1\undefined}
}%
\providecommand \@ifnum [1]{%
 \ifnum #1\expandafter \@firstoftwo
 \else \expandafter \@secondoftwo
 \fi
}%
\providecommand \@ifx [1]{%
 \ifx #1\expandafter \@firstoftwo
 \else \expandafter \@secondoftwo
 \fi
}%
\providecommand \natexlab [1]{#1}%
\providecommand \enquote  [1]{``#1''}%
\providecommand \bibnamefont  [1]{#1}%
\providecommand \bibfnamefont [1]{#1}%
\providecommand \citenamefont [1]{#1}%
\providecommand \href@noop [0]{\@secondoftwo}%
\providecommand \href [0]{\begingroup \@sanitize@url \@href}%
\providecommand \@href[1]{\@@startlink{#1}\@@href}%
\providecommand \@@href[1]{\endgroup#1\@@endlink}%
\providecommand \@sanitize@url [0]{\catcode `\\12\catcode `\$12\catcode
  `\&12\catcode `\#12\catcode `\^12\catcode `\_12\catcode `\%12\relax}%
\providecommand \@@startlink[1]{}%
\providecommand \@@endlink[0]{}%
\providecommand \url  [0]{\begingroup\@sanitize@url \@url }%
\providecommand \@url [1]{\endgroup\@href {#1}{\urlprefix }}%
\providecommand \urlprefix  [0]{URL }%
\providecommand \Eprint [0]{\href }%
\providecommand \doibase [0]{http://dx.doi.org/}%
\providecommand \selectlanguage [0]{\@gobble}%
\providecommand \bibinfo  [0]{\@secondoftwo}%
\providecommand \bibfield  [0]{\@secondoftwo}%
\providecommand \translation [1]{[#1]}%
\providecommand \BibitemOpen [0]{}%
\providecommand \bibitemStop [0]{}%
\providecommand \bibitemNoStop [0]{.\EOS\space}%
\providecommand \EOS [0]{\spacefactor3000\relax}%
\providecommand \BibitemShut  [1]{\csname bibitem#1\endcsname}%
\let\auto@bib@innerbib\@empty
%</preamble>
\bibitem [{\citenamefont {M{\'e}zard}\ \emph {et~al.}(2002)\citenamefont
  {M{\'e}zard}, \citenamefont {Parisi},\ and\ \citenamefont
  {Zecchina}}]{mezard2002analytic}%
  \BibitemOpen
  \bibfield  {author} {\bibinfo {author} {\bibfnamefont {M.}~\bibnamefont
  {M{\'e}zard}}, \bibinfo {author} {\bibfnamefont {G.}~\bibnamefont {Parisi}},
  \ and\ \bibinfo {author} {\bibfnamefont {R.}~\bibnamefont {Zecchina}},\
  }\href@noop {} {\bibfield  {journal} {\bibinfo  {journal} {Science}\ }\textbf
  {\bibinfo {volume} {297}},\ \bibinfo {pages} {812} (\bibinfo {year}
  {2002})}\BibitemShut {NoStop}%
\bibitem [{\citenamefont {Friedgut}\ and\ \citenamefont
  {Bourgain}(1999)}]{friedgut1999sharp}%
  \BibitemOpen
  \bibfield  {author} {\bibinfo {author} {\bibfnamefont {E.}~\bibnamefont
  {Friedgut}}\ and\ \bibinfo {author} {\bibfnamefont {J.}~\bibnamefont
  {Bourgain}},\ }\href@noop {} {\bibfield  {journal} {\bibinfo  {journal}
  {Journal of the American Mathematical Society}\ }\textbf {\bibinfo {volume}
  {12}},\ \bibinfo {pages} {1017} (\bibinfo {year} {1999})}\BibitemShut
  {NoStop}%
\bibitem [{\citenamefont {Kirkpatrick}\ and\ \citenamefont
  {Selman}(1994)}]{kirkpatrick1994critical}%
  \BibitemOpen
  \bibfield  {author} {\bibinfo {author} {\bibfnamefont {S.}~\bibnamefont
  {Kirkpatrick}}\ and\ \bibinfo {author} {\bibfnamefont {B.}~\bibnamefont
  {Selman}},\ }\href@noop {} {\bibfield  {journal} {\bibinfo  {journal}
  {Science}\ }\textbf {\bibinfo {volume} {264}},\ \bibinfo {pages} {1297}
  (\bibinfo {year} {1994})}\BibitemShut {NoStop}%
\bibitem [{\citenamefont {Monasson}\ \emph {et~al.}(1999)\citenamefont
  {Monasson}, \citenamefont {Zecchina}, \citenamefont {Kirkpatrick},
  \citenamefont {Selman},\ and\ \citenamefont
  {Troyansky}}]{monasson1999determining}%
  \BibitemOpen
  \bibfield  {author} {\bibinfo {author} {\bibfnamefont {R.}~\bibnamefont
  {Monasson}}, \bibinfo {author} {\bibfnamefont {R.}~\bibnamefont {Zecchina}},
  \bibinfo {author} {\bibfnamefont {S.}~\bibnamefont {Kirkpatrick}}, \bibinfo
  {author} {\bibfnamefont {B.}~\bibnamefont {Selman}}, \ and\ \bibinfo {author}
  {\bibfnamefont {L.}~\bibnamefont {Troyansky}},\ }\href@noop {} {\bibfield
  {journal} {\bibinfo  {journal} {Nature}\ }\textbf {\bibinfo {volume} {400}},\
  \bibinfo {pages} {133} (\bibinfo {year} {1999})}\BibitemShut {NoStop}%
\bibitem [{\citenamefont {Kempe}\ \emph {et~al.}(2006)\citenamefont {Kempe},
  \citenamefont {Kitaev},\ and\ \citenamefont {Regev}}]{kempe2006complexity}%
  \BibitemOpen
  \bibfield  {author} {\bibinfo {author} {\bibfnamefont {J.}~\bibnamefont
  {Kempe}}, \bibinfo {author} {\bibfnamefont {A.}~\bibnamefont {Kitaev}}, \
  and\ \bibinfo {author} {\bibfnamefont {O.}~\bibnamefont {Regev}},\
  }\href@noop {} {\bibfield  {journal} {\bibinfo  {journal} {SIAM Journal on
  Computing}\ }\textbf {\bibinfo {volume} {35}},\ \bibinfo {pages} {1070}
  (\bibinfo {year} {2006})}\BibitemShut {NoStop}%
\bibitem [{\citenamefont {Bravyi}(2006)}]{bravyi2006efficient}%
  \BibitemOpen
  \bibfield  {author} {\bibinfo {author} {\bibfnamefont {S.}~\bibnamefont
  {Bravyi}},\ }\href@noop {} {\emph {\bibinfo {title} {Efficient algorithm for
  a quantum analogue of {2-SAT}}}},\ \bibinfo {type} {Tech. Rep.}\ (\bibinfo
  {year} {2006})\BibitemShut {NoStop}%
\bibitem [{\citenamefont {Gosset}\ and\ \citenamefont
  {Nagaj}(2013)}]{gosset2013quantum}%
  \BibitemOpen
  \bibfield  {author} {\bibinfo {author} {\bibfnamefont {D.}~\bibnamefont
  {Gosset}}\ and\ \bibinfo {author} {\bibfnamefont {D.}~\bibnamefont {Nagaj}},\
  }\href@noop {} {\bibfield  {journal} {\bibinfo  {journal} {arXiv:1302.0290}\
  } (\bibinfo {year} {2013})}\BibitemShut {NoStop}%
\bibitem [{Note1()}]{Note1}%
  \BibitemOpen
  \bibinfo {note} {To be more precise, $\protect \mathbf {QMA}_1$ is the
  quantum analog of $\protect \mathbf {MA}_1$, where the verification procedure
  allows false instances to be accepted with a small probability.}\BibitemShut
  {Stop}%
\bibitem [{\citenamefont {Laumann}\ \emph
  {et~al.}(2010{\natexlab{a}})\citenamefont {Laumann}, \citenamefont
  {Moessner}, \citenamefont {Scardicchio},\ and\ \citenamefont
  {Sondhi}}]{laumann2010random}%
  \BibitemOpen
  \bibfield  {author} {\bibinfo {author} {\bibfnamefont {C.~R.}\ \bibnamefont
  {Laumann}}, \bibinfo {author} {\bibfnamefont {R.}~\bibnamefont {Moessner}},
  \bibinfo {author} {\bibfnamefont {A.}~\bibnamefont {Scardicchio}}, \ and\
  \bibinfo {author} {\bibfnamefont {S.~L.}\ \bibnamefont {Sondhi}},\
  }\href@noop {} {\bibfield  {journal} {\bibinfo  {journal} {Quantum
  Information \& Computation}\ }\textbf {\bibinfo {volume} {10}},\ \bibinfo
  {pages} {1} (\bibinfo {year} {2010}{\natexlab{a}})}\BibitemShut {NoStop}%
\bibitem [{\citenamefont {Laumann}\ \emph
  {et~al.}(2010{\natexlab{b}})\citenamefont {Laumann}, \citenamefont
  {L{\"a}uchli}, \citenamefont {Moessner}, \citenamefont {Scardicchio},\ and\
  \citenamefont {Sondhi}}]{laumann2010product}%
  \BibitemOpen
  \bibfield  {author} {\bibinfo {author} {\bibfnamefont {C.~R.}\ \bibnamefont
  {Laumann}}, \bibinfo {author} {\bibfnamefont {A.~M.}\ \bibnamefont
  {L{\"a}uchli}}, \bibinfo {author} {\bibfnamefont {R.}~\bibnamefont
  {Moessner}}, \bibinfo {author} {\bibfnamefont {A.}~\bibnamefont
  {Scardicchio}}, \ and\ \bibinfo {author} {\bibfnamefont {S.~L.}\ \bibnamefont
  {Sondhi}},\ }\href@noop {} {\bibfield  {journal} {\bibinfo  {journal}
  {Physical Review A}\ }\textbf {\bibinfo {volume} {81}},\ \bibinfo {pages}
  {062345} (\bibinfo {year} {2010}{\natexlab{b}})}\BibitemShut {NoStop}%
\bibitem [{\citenamefont {Laumann}\ \emph {et~al.}(2012)\citenamefont
  {Laumann}, \citenamefont {Moessner}, \citenamefont {Scardicchio},\ and\
  \citenamefont {Sondhi}}]{laumann2012statistical}%
  \BibitemOpen
  \bibfield  {author} {\bibinfo {author} {\bibfnamefont {C.~R.}\ \bibnamefont
  {Laumann}}, \bibinfo {author} {\bibfnamefont {R.}~\bibnamefont {Moessner}},
  \bibinfo {author} {\bibfnamefont {A.}~\bibnamefont {Scardicchio}}, \ and\
  \bibinfo {author} {\bibfnamefont {S.~L.}\ \bibnamefont {Sondhi}},\ }in\
  \href@noop {} {\emph {\bibinfo {booktitle} {Modern Theories of Many-Particle
  Systems in Condensed Matter Physics}}}\ (\bibinfo  {publisher} {Springer},\
  \bibinfo {year} {2012})\ pp.\ \bibinfo {pages} {295--332}\BibitemShut
  {NoStop}%
\bibitem [{\citenamefont {Ambainis}\ \emph {et~al.}(2012)\citenamefont
  {Ambainis}, \citenamefont {Kempe},\ and\ \citenamefont
  {Sattath}}]{ambainis2012quantum}%
  \BibitemOpen
  \bibfield  {author} {\bibinfo {author} {\bibfnamefont {A.}~\bibnamefont
  {Ambainis}}, \bibinfo {author} {\bibfnamefont {J.}~\bibnamefont {Kempe}}, \
  and\ \bibinfo {author} {\bibfnamefont {O.}~\bibnamefont {Sattath}},\
  }\href@noop {} {\bibfield  {journal} {\bibinfo  {journal} {Journal of the
  ACM}\ }\textbf {\bibinfo {volume} {59}},\ \bibinfo {pages} {24} (\bibinfo
  {year} {2012})}\BibitemShut {NoStop}%
\bibitem [{\citenamefont {Castellana}\ and\ \citenamefont
  {Zdeborov\'{a}}(2011)}]{castellana-zdeborova}%
  \BibitemOpen
  \bibfield  {author} {\bibinfo {author} {\bibfnamefont {M.}~\bibnamefont
  {Castellana}}\ and\ \bibinfo {author} {\bibfnamefont {L.}~\bibnamefont
  {Zdeborov\'{a}}},\ }\href@noop {} {\bibfield  {journal} {\bibinfo  {journal}
  {Journal of Statistical Mechanics: Theory and Experiment}\ }\textbf {\bibinfo
  {volume} {P03023}} (\bibinfo {year} {2011})}\BibitemShut {NoStop}%
\bibitem [{\citenamefont {Bravyi}\ \emph {et~al.}(2009)\citenamefont {Bravyi},
  \citenamefont {Moore},\ and\ \citenamefont {Russell}}]{bravyi2009bounds}%
  \BibitemOpen
  \bibfield  {author} {\bibinfo {author} {\bibfnamefont {S.}~\bibnamefont
  {Bravyi}}, \bibinfo {author} {\bibfnamefont {C.}~\bibnamefont {Moore}}, \
  and\ \bibinfo {author} {\bibfnamefont {A.}~\bibnamefont {Russell}},\
  }\href@noop {} {\bibfield  {journal} {\bibinfo  {journal} {arXiv:0907.1297}\
  } (\bibinfo {year} {2009})}\BibitemShut {NoStop}%
\bibitem [{\citenamefont {Hartmann}\ and\ \citenamefont
  {Weigt}(2005)}]{hartmann-weigt}%
  \BibitemOpen
  \bibfield  {author} {\bibinfo {author} {\bibfnamefont {A.~K.}\ \bibnamefont
  {Hartmann}}\ and\ \bibinfo {author} {\bibfnamefont {M.}~\bibnamefont
  {Weigt}},\ }\href@noop {} {\emph {\bibinfo {title} {Phase transitions in
  combinatorial optimization problems}}}\ (\bibinfo  {publisher} {Whiley-VCH},\
  \bibinfo {address} {Weinheim},\ \bibinfo {year} {2005})\BibitemShut {NoStop}%
\bibitem [{\citenamefont {Arora}\ and\ \citenamefont
  {Barak}(2009)}]{arora-barak}%
  \BibitemOpen
  \bibfield  {author} {\bibinfo {author} {\bibfnamefont {S.}~\bibnamefont
  {Arora}}\ and\ \bibinfo {author} {\bibfnamefont {B.}~\bibnamefont {Barak}},\
  }\href@noop {} {\emph {\bibinfo {title} {Computational complexity. A modern
  approach}}}\ (\bibinfo  {publisher} {Cambridge University Press},\ \bibinfo
  {address} {Cambridge},\ \bibinfo {year} {2009})\BibitemShut {NoStop}%
\bibitem [{Note2()}]{Note2}%
  \BibitemOpen
  \bibinfo {note} {This might appear hopeless. However, AdSAT is a graph
  property, so it is imaginable that it has a compressed testing procedure.
  Moreover, our attempts to show that AdSAT is $\protect \mathbf {\Sigma
  }_2^p$-complete were not successful.}\BibitemShut {Stop}%
\bibitem [{\citenamefont {Davis}\ and\ \citenamefont
  {Putnam}(1960)}]{davis-putnam}%
  \BibitemOpen
  \bibfield  {author} {\bibinfo {author} {\bibfnamefont {M.}~\bibnamefont
  {Davis}}\ and\ \bibinfo {author} {\bibfnamefont {H.}~\bibnamefont {Putnam}},\
  }\href@noop {} {\bibfield  {journal} {\bibinfo  {journal} {Journal of the
  ACM}\ }\textbf {\bibinfo {volume} {7}},\ \bibinfo {pages} {201} (\bibinfo
  {year} {1960})}\BibitemShut {NoStop}%
\bibitem [{\citenamefont {Davis}\ \emph {et~al.}(1962)\citenamefont {Davis},
  \citenamefont {Logemann},\ and\ \citenamefont
  {Loveland}}]{davis-logemann-loveland}%
  \BibitemOpen
  \bibfield  {author} {\bibinfo {author} {\bibfnamefont {M.}~\bibnamefont
  {Davis}}, \bibinfo {author} {\bibfnamefont {G.}~\bibnamefont {Logemann}}, \
  and\ \bibinfo {author} {\bibfnamefont {D.}~\bibnamefont {Loveland}},\
  }\href@noop {} {\bibfield  {journal} {\bibinfo  {journal} {Communications of
  the ACM}\ }\textbf {\bibinfo {volume} {5}},\ \bibinfo {pages} {394} (\bibinfo
  {year} {1962})}\BibitemShut {NoStop}%
\bibitem [{\citenamefont {S\"{o}rensson}\ and\ \citenamefont
  {E\'{e}n}()}]{minisat}%
  \BibitemOpen
  \bibfield  {author} {\bibinfo {author} {\bibfnamefont {N.}~\bibnamefont
  {S\"{o}rensson}}\ and\ \bibinfo {author} {\bibfnamefont {N.}~\bibnamefont
  {E\'{e}n}},\ }\href@noop {} {\enquote {\bibinfo {title} {The \textsc{MiniSat}
  solver},}\ }\bibinfo {howpublished} {\url{http://www.minisat.se}}\BibitemShut
  {NoStop}%
\bibitem [{\citenamefont {Bayardo}\ and\ \citenamefont
  {Pehousek}(2000)}]{relsat}%
  \BibitemOpen
  \bibfield  {author} {\bibinfo {author} {\bibfnamefont {R.~J.~J.}\
  \bibnamefont {Bayardo}}\ and\ \bibinfo {author} {\bibfnamefont {J.~D.}\
  \bibnamefont {Pehousek}},\ }in\ \href@noop {} {\emph {\bibinfo {booktitle}
  {Proceedings of the 17th AAAI}}}\ (\bibinfo {year} {2000})\ pp.\ \bibinfo
  {pages} {157--162}\BibitemShut {NoStop}%
\bibitem [{\citenamefont {Bayardo}()}]{relsat-web}%
  \BibitemOpen
  \bibfield  {author} {\bibinfo {author} {\bibfnamefont {R.~J.~J.}\
  \bibnamefont {Bayardo}},\ }\href@noop {} {\enquote {\bibinfo {title} {Relsat.
  a propositional satisfiability solver and model counter},}\ }\bibinfo
  {howpublished} {\url{https://code.google.com/p/relsat/}}\BibitemShut
  {NoStop}%
\bibitem [{\citenamefont {Kirkpatrick}\ \emph {et~al.}(1983)\citenamefont
  {Kirkpatrick}, \citenamefont {Gelatt},\ and\ \citenamefont
  {Vecchi}}]{kirkpatrick-gelatt-vecchi}%
  \BibitemOpen
  \bibfield  {author} {\bibinfo {author} {\bibfnamefont {S.}~\bibnamefont
  {Kirkpatrick}}, \bibinfo {author} {\bibfnamefont {D.~C.}\ \bibnamefont
  {Gelatt}}, \ and\ \bibinfo {author} {\bibfnamefont {M.~P.}\ \bibnamefont
  {Vecchi}},\ }\href@noop {} {\bibfield  {journal} {\bibinfo  {journal}
  {Science}\ }\textbf {\bibinfo {volume} {220}},\ \bibinfo {pages} {671}
  (\bibinfo {year} {1983})}\BibitemShut {NoStop}%
\bibitem [{\citenamefont {Papadimitriou}(1991)}]{papadimitriou}%
  \BibitemOpen
  \bibfield  {author} {\bibinfo {author} {\bibfnamefont {C.~H.}\ \bibnamefont
  {Papadimitriou}},\ }in\ \href@noop {} {\emph {\bibinfo {booktitle}
  {Proceedings of the 32nd Annual IEEE Symposium on Foundations of Computer
  Science}}}\ (\bibinfo {year} {1991})\ pp.\ \bibinfo {pages}
  {163--169}\BibitemShut {NoStop}%
\bibitem [{\citenamefont {Sch\"{o}ning}(1999)}]{schoening-1}%
  \BibitemOpen
  \bibfield  {author} {\bibinfo {author} {\bibfnamefont {U.}~\bibnamefont
  {Sch\"{o}ning}},\ }in\ \href@noop {} {\emph {\bibinfo {booktitle}
  {Proceedings of the 40th Annual IEEE Symposium on Foundations of Computer
  Science}}}\ (\bibinfo {year} {1999})\ pp.\ \bibinfo {pages}
  {410--414}\BibitemShut {NoStop}%
\bibitem [{\citenamefont {Sch\"{o}ning}(2002)}]{schoening-2}%
  \BibitemOpen
  \bibfield  {author} {\bibinfo {author} {\bibfnamefont {U.}~\bibnamefont
  {Sch\"{o}ning}},\ }\href@noop {} {\bibfield  {journal} {\bibinfo  {journal}
  {Algorithmica}\ }\textbf {\bibinfo {volume} {32}},\ \bibinfo {pages} {615}
  (\bibinfo {year} {2002})}\BibitemShut {NoStop}%
\bibitem [{Note3()}]{Note3}%
  \BibitemOpen
  \bibinfo {note} {In practice it is determined in the following way: we
  compute $|\Phi _{\protect \mathrm {UNSAT}} - 1/2|$ for all the values of
  $\alpha $, rank them in ascending order and select the first five values;
  then we make a linear interpolation between these points and use the
  interpolating function to compute $\alpha ^*$, the value of $\alpha $
  corresponding to $\Phi _{\protect \mathrm {UNSAT}} = 1/2$.}\BibitemShut
  {Stop}%
\bibitem [{Note4()}]{Note4}%
  \BibitemOpen
  \bibinfo {note} {Being optimistic, one could conjecture that the complexity
  of the problem is $\gamma \sim 1$ as seen for the low $N$ numerics. This
  means that, assuming a sufficient number of iterations (probably in the
  millions), the number of restarts necessary to get a converged $\alpha _N^*$
  for $N=150$ is about $2^{150}\sim 10^{45}$.}\BibitemShut {Stop}%
\bibitem [{\citenamefont {Zhang}\ \emph {et~al.}(2012)\citenamefont {Zhang},
  \citenamefont {Ramezanpour}, \citenamefont {Zdeborov{\'a}},\ and\
  \citenamefont {Zecchina}}]{zhang2012message}%
  \BibitemOpen
  \bibfield  {author} {\bibinfo {author} {\bibfnamefont {P.}~\bibnamefont
  {Zhang}}, \bibinfo {author} {\bibfnamefont {A.}~\bibnamefont {Ramezanpour}},
  \bibinfo {author} {\bibfnamefont {L.}~\bibnamefont {Zdeborov{\'a}}}, \ and\
  \bibinfo {author} {\bibfnamefont {R.}~\bibnamefont {Zecchina}},\ }\href@noop
  {} {\bibfield  {journal} {\bibinfo  {journal} {Journal of Statistical
  Mechanics: Theory and Experiment}\ }\textbf {\bibinfo {volume} {P05025}}
  (\bibinfo {year} {2012})}\BibitemShut {NoStop}%
\end{thebibliography}%

\end{document}